\documentclass[twocolumn]{aastex631}
\usepackage{amsmath}
\usepackage{comment}
\usepackage{gensymb}
\usepackage{bm}
\usepackage{xspace}
\usepackage{ctable}

\usepackage{algorithm2e}
\usepackage{algorithmic}
\usepackage{array}
\usepackage{graphicx}

\def\Ncal{\mathcal{N}}

\def\Ucal{\mathcal{U}}

\def\SlewTime{\mathtt{SlewTime}}

\def\RelRed{\mathtt{RelRed}}
\def\real{\mathtt{real}}
\def\NumFeas{\mathtt{NumFeas}}
\def\NumOptimal{\mathtt{NumOptimal}}
\def\Runtime{\mathtt{Runtime}}

\def\TTPGLOBAL{\textsf{TTP-Global}\xspace}
\def\LSONE{\textsf{TTP-LS-1}\xspace}
\def\LSTEN{\textsf{TTP-LS-10}\xspace}

\def\slew{\mathrm{slew}}
\def\by{\mathrm{by}}
\def\at{\mathrm{at}}

\def\slote{\mathtt{slot,e}}
\def\slotell{\mathtt{slot,\ell}}

\begin{document}

\title{Solving the Traveling Telescope Problem with Mixed Integer Linear Programming}

\author[0000-0002-9305-5101]{Luke B. Handley}
\affiliation{Department of Physics \& Astronomy, University of California Los Angeles, Los Angeles, CA 90095, USA}

\author[0000-0003-0967-2893]{Erik A.\ Petigura}
\affiliation{Department of Physics \& Astronomy, University of California Los Angeles, Los Angeles, CA 90095, USA}

\author[0000-0002-8952-5617]{Velibor V. Mi\v{s}i\'{c}} \affiliation{Decisions, Operations and Technology Management, Anderson School of Management, University of California Los Angeles, Los Angeles, CA 90095, USA}



\begin{abstract}
The size and complexity of modern astronomical surveys has grown to the point where, in many cases, traditional human scheduling of observations are tedious at best and impractical at worst. Automated scheduling algorithms present an opportunity to save human effort and increase scientific productivity. A common scheduling challenge involves determining the optimal ordering of a set of targets over a night subject to timing constraints and time-dependent slew overheads. We present a solution to the `Traveling Telescope Problem' (TTP) that uses Mixed-Integer Linear Programming (MILP). This algorithm is fast enough to enable dynamic schedule generation in many astronomical contexts. It can determine the optimal solution for 100 observations within 10 minutes on a modern workstation, reducing slew overheads by a factor of 5 compared to random ordering. We also provide a heuristic method that can return a near-optimal solution at significantly reduced computational cost. As a case study, we explore our  algorithm's suitability to automatic schedule generation for Doppler planet searches.
\end{abstract}



\section{Introduction} 
\label{sec:intro}

Maximizing the scientific yield of expensive and often oversubscribed astronomical instrumentation requires meticulous planning of each night's observations. However, determining the optimal (or even near-optimal) sequence of observations is challenging and time consuming task. Schedulers must incorporate temporal accessibility windows while factoring in slew and acquisition overheads that can themselves be time-variable. In this paper, we refer to the task of determining the optimal ordering of a set of observations by a telescope as the `Traveling Telescope Problem' or `TTP' given its similarities to the `Traveling Salesman Problem' or `TSP'.

The scientific benefits of intelligently sequenced observations can be significant, especially for programs with many targets spread over the entire sky. As an example, the Doppler planet searches at the Keck-I telescope observe up to 100 targets per night \citep{Howard10}. As we show below, a quasi-random sequence of 100 targets requires over 3 hours of slew during a 10 hour night while an optimized sequence can reduce this to 0.5 hours. 

Automated solutions to the TTP offer a number of opportunities. As is the case for the TSP, a skilled human scheduler can generate a observation sequence that significantly outperforms a random ordering. However, such script generation takes significant human effort that could be devoted elsewhere. In addition, a number of ongoing and forthcoming surveys are or will be scheduled automatically. The Zwicky Transient Facility (ZTF; \citealt{ZTFoverview}) and the Legacy Survey of Space and Time (LSST; \citealt{LSST}) are two examples. Automated solutions to the TTP are necessary for automated surveys. 

\begin{figure*}
    \centering
    \includegraphics[width=\linewidth]{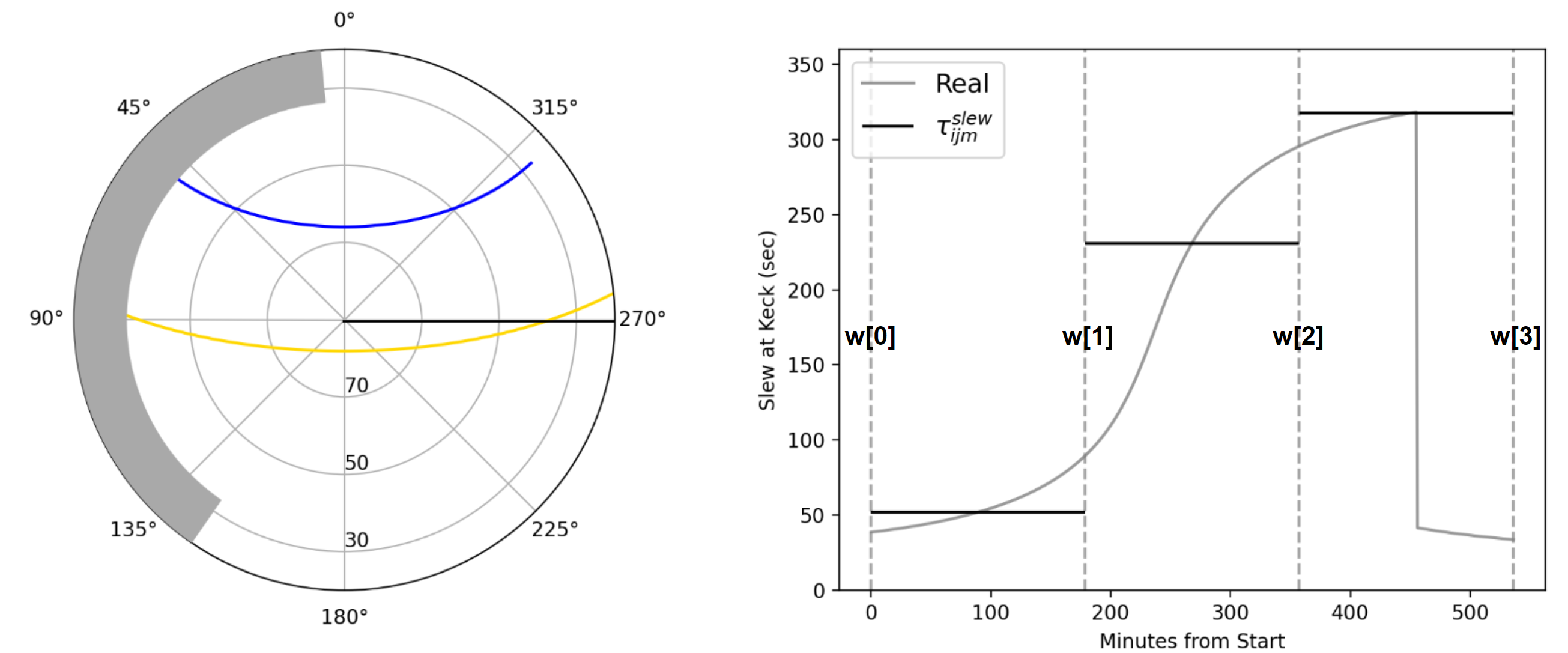}
    \caption{\textbf{Time dependent slew overheads at Keck Observatory.} The left plot shows the motion of two targets in the alt/az frame across the accessible region at Keck Observatory (latitude = $19.8\degree$ N). The top target sits at $\delta = 43.8\degree$, the bottom at $\delta=11.7\degree$. As time progresses, the azimuthal separation dominates the slew and steadily increases to a maximum of over five minutes. Late in the night, the lower declination target crosses azimuth = $270\degree$, which corresponds to the assumed telescope cable wrap limit. At this time, a direct slew becomes available, and the slew time drops significantly.
    For any two arbitrary targets $i$ and $j$, $\tau^\mathrm{slew}_{ijm}$ approximates the pairwise slew time within the bounds of each time slot as dictated by $M$. The right plot shows the interpolated slew curve (grey) with the travel time tensor $\tau^\mathrm{slew}_{ijm}$ (black) over-plotted for the full night with $M=3$. In this extreme case, $\tau^\mathrm{slew}_{ijm}$ often misrepresents the slew time by several minutes in the second and third time slot.}

    \label{fig:tauslew}
\end{figure*}

While a rich literature exists on the TSP, standard solutions do not directly transfer to the TTP for several reasons. First, targets are often only accessible for a fraction of the night either due to their position on the sky or scientific need for time-critical observations. Thus, a large fraction of the $N!$ target sequences are infeasible. Second, slew time between targets is {\em itself} a function of time. As an example, Figure~\ref{fig:tauslew} shows the trajectory of two targets above the Keck-I telescope atop Maunakea. Two targets cross the meridian north and south of zenith, respectively. Like most large telescopes, Keck-I moves in the altitude and azimuth directions and target slews are dominated by differences in azimuth. Figure~\ref{fig:tauslew} shows the variation in slew time over the course of the night, which grows as the targets first approach the meridian and then straddle the telescopes cable wrap limits.

Previous efforts have addressed the TTP under a number of simplifying assumptions. The ZTF scheduler divides the night into short intervals and solves a standard TSP while treating the set of accessible targets and target-to-target slew times as constant with in the interval \citep{ZTFscheduler}. The {\em James Webb Space Telescope's} scheduling architecture \citep{jwstscheduler}, which is built upon the {\em Hubble Space Telescope's} SPIKE software \citep{spike}, also treats target-to-target slew times as constant. At present, we are not aware of any global solutions to the TTP that capture both variable accessibility windows and slew overheads.

This paper is organized as follows. We describe the TTP in Section \ref{sec:ttp} and present a formulation using Mixed-Integer Linear Programming that can be solved to global optimally range of problem sizes. In Section \ref{sec:performance} we conduct a  suite numerical experiments to determine the performance and computational cost of our algorithm over various problem sizes. Section \ref{sec:discussion} discusses the limitations of our global approach and the potential of heuristic solutions. We conclude in Section \ref{sec:conclusion}.

\section{Formulation of Traveling Telescope Problem} 
\label{sec:ttp}

\subsection{Problem Setup}
For a given set of targets and an observing interval, we seek the tour that completes all exposures in the shortest possible time. Targets must be accessible for at least part of the observing duration. The slew time between any  pair of targets must be computed in advance, but the slew time may {\em itself} be a function of time. The formulation presented below was inspired by \cite{tw} who developed a framework to optimize the profitability of parcel pickup and delivery with variable time windows and travel times.

\subsection{Slot Framework} \label{sec:slots}

Following the TSP literature, we refer to targets to be traversed in the TTP as \textit{nodes} since that work emphasizes that the solution is a directed graph connecting all targets. With a list of $N$ targets to be scheduled, we define the total set of nodes to be $\{ 0,1,...,N,N+1\}$. The nodes $0$ and $N+1$ are the anchoring start and end nodes; their location is arbitrary, i.e. not associated with a celestial source. Their purpose is explained in Section \ref{sec:nightcon}. Each node $i$ has an associated exposure time $\tau^\mathrm{exp}_{i}$, and accessibility window  $[t_{i}^\mathrm{e}, t_{i}^\mathrm{\ell}]$. The values $t_{i}^\mathrm{e}$ and $t_{i}^\mathrm{\ell}$ are the earliest and latest times the tour can depart node $i$, i.e. the time the observation concludes (see Figure~\ref{fig:windows}). We summarize the symbols from the main body of this text in Appendix~\ref{sec:variables}.

Next, we break the scheduling interval into $M$ sub-intervals or `slots', within which travel time is treated as constant. $M$ is a free parameter and impacts the computational load (see Section \ref{sec:performance}). The slots may have uneven durations if desired. The boundaries of the slots are $[w_{m},w_{m+1}]$ where $m$ indexes each slot.

We then construct the slew tensor $\tau^\mathrm{slew}_{ijm}$ that specifies the travel time between any two nodes during every slot (see Figure \ref{fig:tauslew}). This depends on the telescope slew speed in altitude and azimuth directions as well as cable-wrap considerations. For a concrete example, $\tau^\mathrm{slew}_{4,6,2}$ specifies the computed travel time between the $i=4$ node and $j=6$ node during the bounds of the slot $m=2$.

\begin{figure*}
\centering
\includegraphics[width=\linewidth]{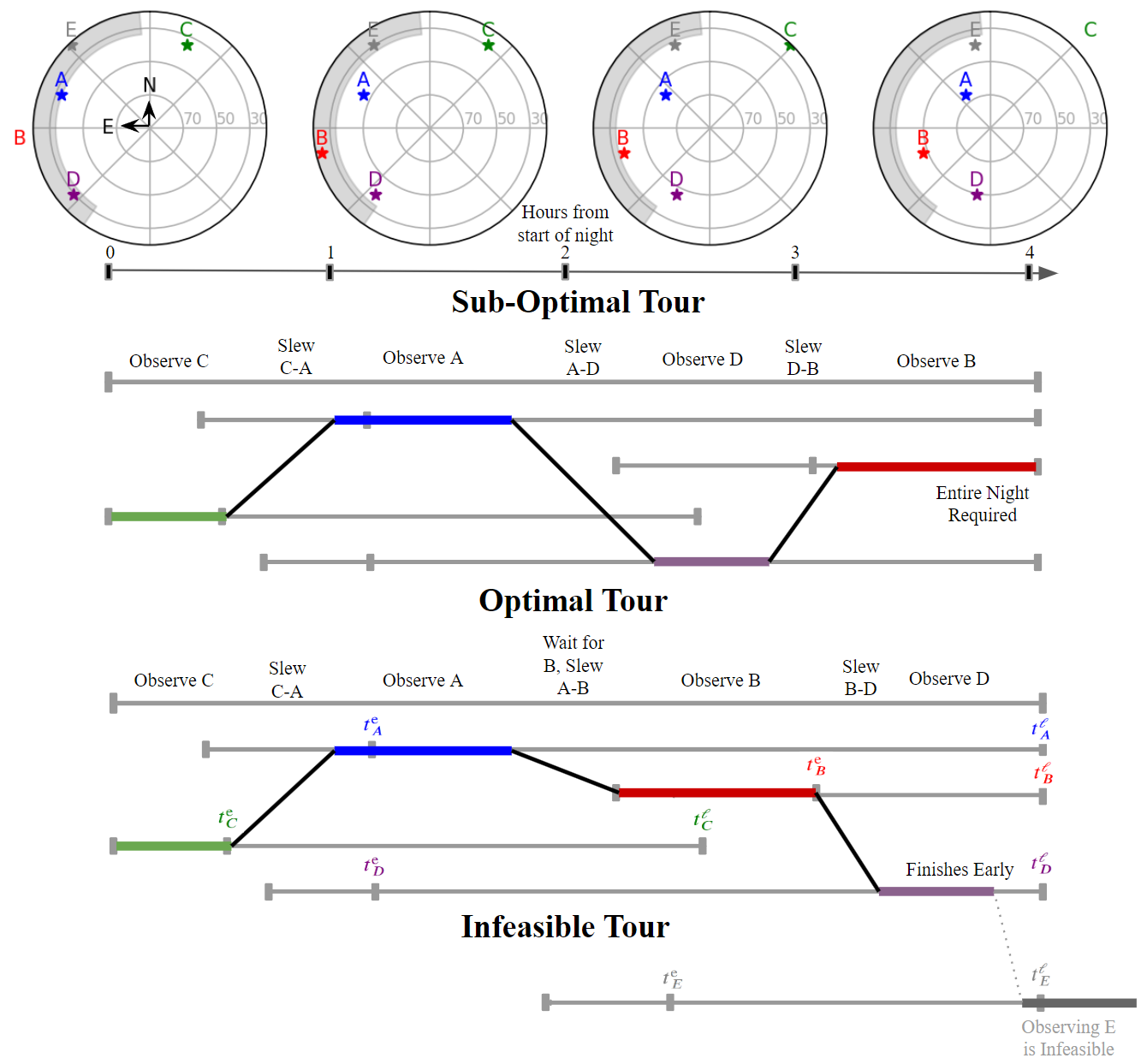}
\caption{{\bf Schematic of sub-optimal, optimal, and infeasible tours in the travelling telescope problem}. The four polar plots on top show the sky above Keck Observatory (latitude = $+19$~deg) in alt/az coordinates. During a duration of four hours, targets A--E move across the sky as the Earth rotates. The shaded region shows an inaccessible region of alt/az (here, the Keck-I Naysmth platform). 
We have exaggerated slew times to clearly illustrate the differences between the three tours. In the top {\bf sub-optimal tour}, we include a horizontal timeline for targets A--D. The windows of accessibility are shown as vertical ticks. For example, the earliest the telescope can {\em complete} observations of target A is labeled with $t_A^{e}$ while $t_A^{l}$, denotes the latest time. Note that $t_A^{e}$ depends on both the rise time and exposure duration, and $t_{i}^\mathrm{\ell}$ is the either the set time or the end of the observing interval, whichever comes first.
This scheduler uses a greedy approach; i.e., once the observation A of completes, the telescope immediately slews to the unobserved target with the shortest slew time. The tour is feasible, but the entire observing duration is required in this example. 
The middle {\bf optimal tour} is scheduled with a global optimization. In our example, the slew between A and D is so long that it is advantageous for the telescope to wait until target B rises to observe it, before proceeding to D. This is the optimal tour to observe targets A--D. Finally the bottom plot shows the inclusion of a fifth target E to illustrate an {\bf infeasible tour}. There is not enough time to observe all targets.}
\label{fig:windows}
\end{figure*}

\subsection{Decision Variables} \label{sec:nightvar}

Our MILP formulation involves both binary and continuous variables. The following variables trace the flow through the nodes and the times of node departures:

\begin{itemize}
\item $X_{ijm}$ is a binary variable equal to 1 if the tour traverses the arc from node $i$ to node $j$ during slot $m$ and 0 otherwise.

\item $Y_{i}$ is a binary variable equal to 1 if the node $i$ is entered at some point during the tour and 0 otherwise.

\item $t_{i}$ is a continuous variable denoting the departure time from node $i$.

\item $t_{ijm}$ is a continuous variable and equal to $t_{i}$ if the tour departs the node $i$ toward the node $j$ during the time slot $m$, 0 otherwise.
\end{itemize}

The non-zero elements of $X_{ijm}$ describe the tour. The tensor dot product of $X_{ijm}$ and $\tau^\mathrm{slew}_{ijm}$

\begin{equation}
    \sum_{i,j = 1}^{N}  \sum_{m = 0}^{M-1} \tau_{ijm}^\mathrm{slew}X_{ijm}
\end{equation}

\noindent is equal to the total travel time.

\subsection{Constraints} \label{sec:nightcon}
Next, we introduce the following constraints:

\vspace{0.2cm}
\noindent\textbf{Constraint 1. Tour must depart the starting node.} We require that flow be non-zero from $0$ to some arbitrary target node $j$ during some slot $m$ in time to begin the tour. 

\begin{equation} \label{eq:origin}
    \sum_{j = 1}^{N} \sum_{m = 0}^{M-1} X_{0,j,m} = 1
\end{equation} 

\noindent\textbf{Constraint 2: Tour must conclude at the ending node}. Similarly, we anchor the end of the tour with the end node $N+1$. We must traverse the arc $(i,N+1)$ from some arbitrary target node $i$. 

\begin{equation} \label{eq:destination}
    \sum_{i = 1}^{N} \sum_{m = 0}^{M-1} X_{i,N+1,m} = 1
\end{equation}

\noindent\textbf{Constraint 3: $Y$ must indicate the visitation of a node}. In the simplest possible case, we have a trivial solution traversing from $0$ to $N+1$, stopping at a single target node along the way. The objective (see Section \ref{sec:nightobj}) will reward the visitation of additional nodes between these two constrained anchors. This requires first defining the variable $Y$ to track whether nodes are being visited. $Y_{i}$ will be activated if the tour flows from the node $i$ during any slot in time, into any subsequent node $j$.

\begin{equation} \label{eq:defy}
    \sum_{j = 1}^{N+1} \sum_{m = 0}^{M-1} X_{ijm} = Y_{i} \quad \forall\, i = 0,\ldots,N 
    \end{equation}

\noindent\textbf{Constraint 4: A slew into any target node must be accompanied by a slew away from that node}. Excluding the starting and ending nodes, we require that the slew into any target node $k$ must be accompanied by a slew out of $k$.
\begin{equation} \label{eq:flow}
    \sum_{i = 0}^{N} \sum_{m = 0}^{M-1} X_{ikm} - \sum_{j = 1}^{N+1} \sum_{m =0}^{M-1} X_{kjm} = 0 \quad \forall\, k = 1,\ldots,N
\end{equation}
If $X_{ijm}$ is 1 for any arc $(i,k)$ in the sequence, it will also hold 1 for some $(k,j)$ arc at an arbitrary time. If the left term is 0 (the tour never enters node $k$) then the tour will never traverse a departing arc originating at $k$. The chronology of these events will next be enforced using the time variable $t$. 

\noindent\textbf{Constraint 5: Define the selection variable} $\bm{t_{ijm}}$. Now we enforce the proper constraints to define the selection time variable $t_{ijm}$ relative to our generic time variable $t_{i}$. By summing over all potential destinations $j$ and time slots $m$, we recover the value stored in $t_{i}$. 

\begin{equation} \label{eq:tdef}
    t_{i} = \sum_{j = 1}^{N+1} \sum_{m = 0}^{M-1} t_{ijm} \quad \forall\, i = 0,\dots,N
\end{equation}
The second time variable $t_{ijm}$ is critical for the following two constraints. It behaves like the product of $t_{i}$ and $X_{ijm}$, but works within a linear program.

\noindent\textbf{Constraint 6: Departure time from a node is greater than the departure from the previous node plus the slew and exposure time}. Say we begin to traverse from one node to another on the arc $(i,j)$ within the bounds of the slot $m$. Before the telescope may depart from the next node $j$, a minimum amount of time must pass, equal to the slew experienced on the journey from $i$ to $j$  plus the exposure time at the new node, i.e. $\tau^\mathrm{slew}_{ijm}$ +  $\tau_{j}^\mathrm{exp}$. The minimum time value of our departure from $j$ is the time that the previous node $i$ was departed $t_{ijm}$ plus this minimum `passing time'. The inclusion of the variable $X_{ijm}$ will ensure only the proper values of $i$ and $m$ are considered for the journey to the new node $j$.

\begin{equation}
\begin{aligned} 
\label{eq:timeconstr}
t_{j} \ge \sum_{i=0}^{N} \sum_{m=0}^{M-1} \left( t_{ijm} + (\tau_{ijm}^\mathrm{slew} + \tau_{j}^\mathrm{exp}) X_{ijm}\right)  
\\ \forall\, j = 1,\ldots,N+1 
\end{aligned}
\end{equation}

\noindent\textbf{Constraint 7: Ensure departure times are consistent with slot bounds}. We must enforce constraints on the departure times $t_{ijm}$ using bounds of the time windows defined in $w$. If we exit the node $i$ during the time slot $m$, then the departure time must be within the bounds of the slot window.

\begin{equation}
\begin{aligned}
w_{m} X_{ijm} \le t_{ijm} \le w_{m+1} X_{ijm} 
\\ \forall\, i = 0,\ldots,N+1 
\\ \forall\, j = 0,\ldots,N+1 
\\ \forall\, m = 0,\ldots,M-1 &\label{eq:tmin}
\end{aligned}
\end{equation}

\noindent\textbf{Constraint 8: Departure times must respect node accessibility.} Finally, we enforce the time window constraints on the departure times $t_{i}$ for all the visited target nodes.

\begin{equation} \label{eq:windows}
    t_{i}^\mathrm{e}Y_{i} \le t_{i} \le t_{i}^\mathrm{\ell}Y_{i} \quad \forall\, i = 1,\ldots,N
\end{equation}

\subsection{Objective} \label{sec:nightobj}

\noindent\textbf{We seek to maximize the the number of scheduled exposures while minimizing total slew time}

\begin{equation} \label{eq:obj}
    \text{Max}\left( \sum_{i =1}^{N} Y_{i} - C\sum_{i,j = 1}^{N}  \sum_{m = 0}^{M-1} \tau_{ijm}^\mathrm{slew}X_{ijm})\right)
\end{equation}
Here, $C$ is a small constant such that the slew penalisation (second term) is always less than unity. Notice that our anchor nodes do not contribute to the total slews with this summation convention, and are therefore used only for constraining the flow of the tour (a physical location need not be set, and the values in the first row and column of $\tau_{ijm}^\mathrm{slew}$ are set to 0).

The objective function does not require all nodes be visited. In similar TSP literature such as \cite{tw}, each target node may be assigned a scalar priority $p_{j}$ included as a coefficient in the left summation term in Equation \ref{eq:obj}. In such a formulation, lowest priority targets are preferentially excluded when total completion is infeasible. We note that 
global optimality becomes less intuitive when targets have different numerical priorities.

In the TTP, the optimization will work to remove the targets that contribute \textit{most} to the total slews in every case. The objective is clear: observe as many targets as possible as quickly as possible. We note that the formulation above describes a simple TTP where targets are observed once and no additional constraints are placed on the timing between observations. Appendix~\ref{sec:variants} describes small modifications to the algorithm presented above that can accommodate such constraints.

\begin{figure*}
    \centering
    \includegraphics[width=\linewidth]{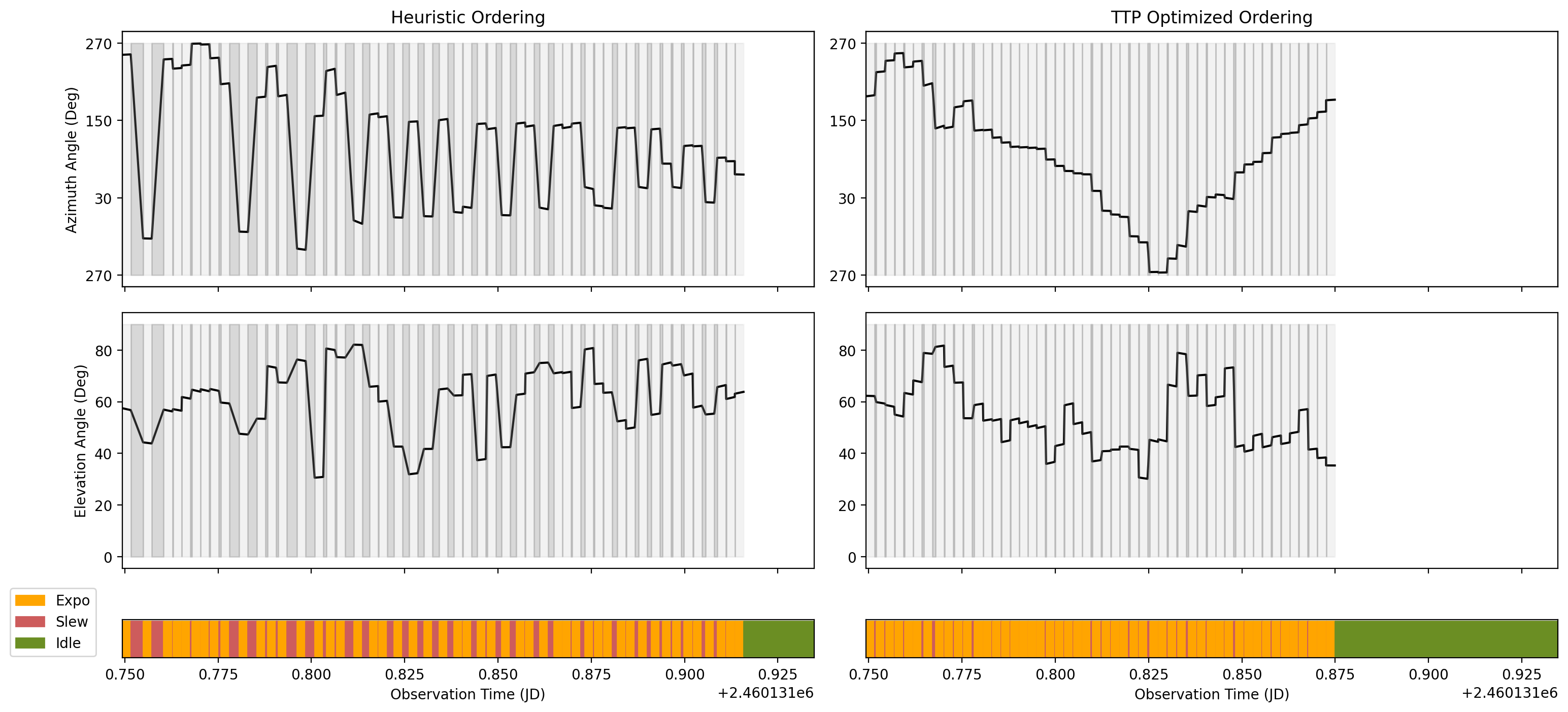}
    \caption{Comparison of a simple heuristic target ordering and a TTP optimized tour of $N=50$ targets during a simulated half night. The former routine orders the targets by their increasing set times. Top row: the azimuthal coordinate of the simulated telescope in each case. Middle row: the elevation angle of the simulated telescope. Bottom row: time spent exposing, slewing, or idle. The targets have randomized time windows, but are all accessible for at least half of the scheduling duration. The worst case slew estimation from Equation \ref{eq:exp} is 3.34 minutes per target. For the left (heuristic) column, the total slew time is 73 minutes, or an average of 91 seconds per slew, leaving 27 minutes idle. In the optimized (right) tour, the total time is a mere 13.8 minutes, or an average of 17.3 seconds per slew, resulting in 86 minutes idle (23 additional targets at the same pace).}
    \label{fig:tourdiagram}
\end{figure*}

\begin{comment}
Random:
Real Total = 72.7 minutes
Real Avg = 90.9 seconds
Optimal:
Real Total = 13.8 minutes
Real Avg = 17.3 seconds
\end{comment}

% method_LS_writeup.tex

% To do :

\section{Performance} \label{sec:performance}

\subsection{Numerical Experiments}

We evaluated our algorithm's ability to solve the TTP through a suite of simulated target lists. We explored problem complexity along the following three axes: number of targets $N$, number of time slots $M$, and duration of the scheduling interval $D$. We designed TTPs in the following manner:

\begin{enumerate}
\item We specified a random calendar night at Keck Observatory.
\item We selected an observing duration $D$.
\item We selected $N$ stars from the California Legacy Survey (CLS; \citealt{CLS}), a collection of 719 nearby stars that have been observed from Keck observatory for several decades as part of an extrasolar planet search. These stars comprise a good TTP test set because they are nearly uniformly distributed on the sky with declination $\delta \gtrsim -40$~deg and are thus observable for $\gtrsim 7$ months per year. When sampling the targets, we required they be accessible for more than 50\% of duration $D$. We removed a few close binaries that have negliable slew speeds.
\item We modeled the slew rate of the Keck telescope as 1~deg per second in both altitude and azimuth directions. 
\item We assigned uniform exposure times (in minutes) to all targets according to:
\begin{equation} \label{eq:exp}
\tau^\mathrm{exp}_{i} = (D - 2 N) / N \quad \forall\, i=1,\ldots,N.
\end{equation}
Setting the exposure times this way would completely fill the observing duration given a random ordering of targets with an average slew time of 120~sec or 2~min. For our experiments, slews in azimuth (which ranges from 0 to 360 deg) dominate over those in altitude (which ranges from 0 to 90 deg). The average distance between two randomly selected values between 0 and 360 is 120. We expect all solutions to the TTP to be significantly more efficient. Thus our experiments are feasible by construction and we report the improvement relative to this random ordering.
\item We selected $M$ uniform slots within $D$.
\item We calculated the distance tensor $\tau^\mathrm{slew}_{ijm}$ at the midpoint of each slot.
\end{enumerate}

With the experiment specified, we solved the MILP described in Section~\ref{sec:ttp} with one additional constraint.

\noindent\textbf{Constraint 9: All target nodes must be visited.}

\begin{equation}
    Y_{i} = 1 \quad \forall\, i = 1,\ldots,N \label{eq:Y_eq_1}
\end{equation}

\noindent

The problem generation process described above in general results in TTP instances in which it is possible to observe all $N$ targets. Given this, we modify the objective function described in equation~\eqref{eq:obj} to focus on minimizing slews only, resulting in the following new objective function.
\begin{equation}
    \text{Min}\left(\sum_{i,j = 1}^{N}  \sum_{m = 0}^{M-1} \tau_{ijm}^\mathrm{slew}X_{ijm}\right) \label{eq:objective_slews_only}
\end{equation}

In our results in Section~\ref{subsec:results_TTPGLOBAL}, we use \TTPGLOBAL to refer to the formulation defined in Section~\ref{sec:ttp} with the objective function in equation~\eqref{eq:objective_slews_only} and constraint~\eqref{eq:Y_eq_1}.

%
\begin{comment}
Even with our worst case slew estimation, one can contrive edge cases where our stochastic sampling may still generate infeasible problems. If every target happened to set near the beginning of the night, it would be impossible to achieve full completion while respecting Equation \ref{eq:windows}. Only fully feasible problems are included in our analysis.
\end{comment}

Before describing our grid-based exploration of problem complexity, we show one example solution to the TTP in Figure~\ref{fig:tourdiagram}. We compared it with a simple heuristic solution for a 50 target observing sequence conducted over a half night to emulate a human generated script. In this heuristic, targets are observed in the order of their set times, i.e. the earliest setting target is observed first. Figure~\ref{fig:tourdiagram} shows graphically the slew overheads that \TTPGLOBAL eliminates through a more efficient ordering.

\subsection{Computational Results}
\label{subsec:results_TTPGLOBAL}
% We run a unique model on each problem configuration with varying values of $M$ to clearly show how the number of time slots affects run time and the accuracy of slew estimations. In other words, for each interval type $D$ and given value of $N$, we test models with varying $M$ on the same set of targets.

We test our formulation using different combinations of the number of targets $N$, the number of slots $M$ and the duration/scheduling interval $D$. For the scheduling interval $D$, we consider quarter nights, half nights and full nights. For quarter nights, we vary $N$ in $\{5,10,25\}$; for half nights, we vary $N$ in $\{5,10,25,50\}$; and for full nights, we vary $N$ in $\{5,10,25,50,100\}$. For each combination of $D$ and $N$, we vary $M$ in $\{1,3,10\}$.

For each combination of $N$ and $D$, we consider 10 randomly generated sets of targets, and consider the three different values of $M$, giving rise to a total of $(3 + 4 + 5) \times 3 \times 10 = 360$ problem instances. We solved \TTPGLOBAL using Gurobi version 10.0.1, a state-of-the-art optimization suite that solves MILP problems using the branch-and-bound algorithm \citep{gurobi}. We used the Python programming language to generate the input data for \TTPGLOBAL and to formulate \TTPGLOBAL using the Gurobi Python API. We conducted our suite of numerical experiments on Amazon Elastic Compute Cloud (EC2), on a single instance of type {\tt m6a.48xlarge} (AMD EPYC 7R13 processor, with 192 virtual CPUs and 768 GB of memory). For each experiment, we allocated 8 virtual CPUs and limited computation time to 1800 seconds. 

For a few points of reference, a TTP with $N=25$ targets and $M=1$ involved an MILP with 1643 rows (constraints) and 1513 columns (variables), and the $M=10$ case had 14765 rows and 14635 columns. A TTP with $N=100$ targets had 21518 rows, 21013 columns for $M=1$, and 208790 rows, 208285 columns for $M=10$.

For each experiment, we recorded the following information: 
\begin{itemize}
\item $\SlewTime_{\tau}$: total slew time of the final schedule obtained from Gurobi, calculated using the discretized tensor $\tau^\mathrm{slew}_{ijm}$. 
\item $\RelRed_{\tau}$: relative reduction in slew time of the final schedule compared to the randomly ordered value of $2N$; mathematically, it is defined as:
\begin{equation} \label{eq:relredtau}
\RelRed_{\tau} = \frac{2N - \SlewTime_{\tau}}{2N} \times 100\%.
\end{equation}
\item $\SlewTime_{\real}$: real slew time of the final schedule, calculated based on the $t_i$ departure time values of the schedule.
\item $\RelRed_{\real}$: the analog of $\RelRed_{\real}$ for the real slew time.
\item $\Runtime$: computation time
\item Whether a provably optimal solution was found.
\item Whether a feasible solution was found.
\end{itemize}

\begin{figure*}
    \centering
    \includegraphics[width=0.9\textwidth]{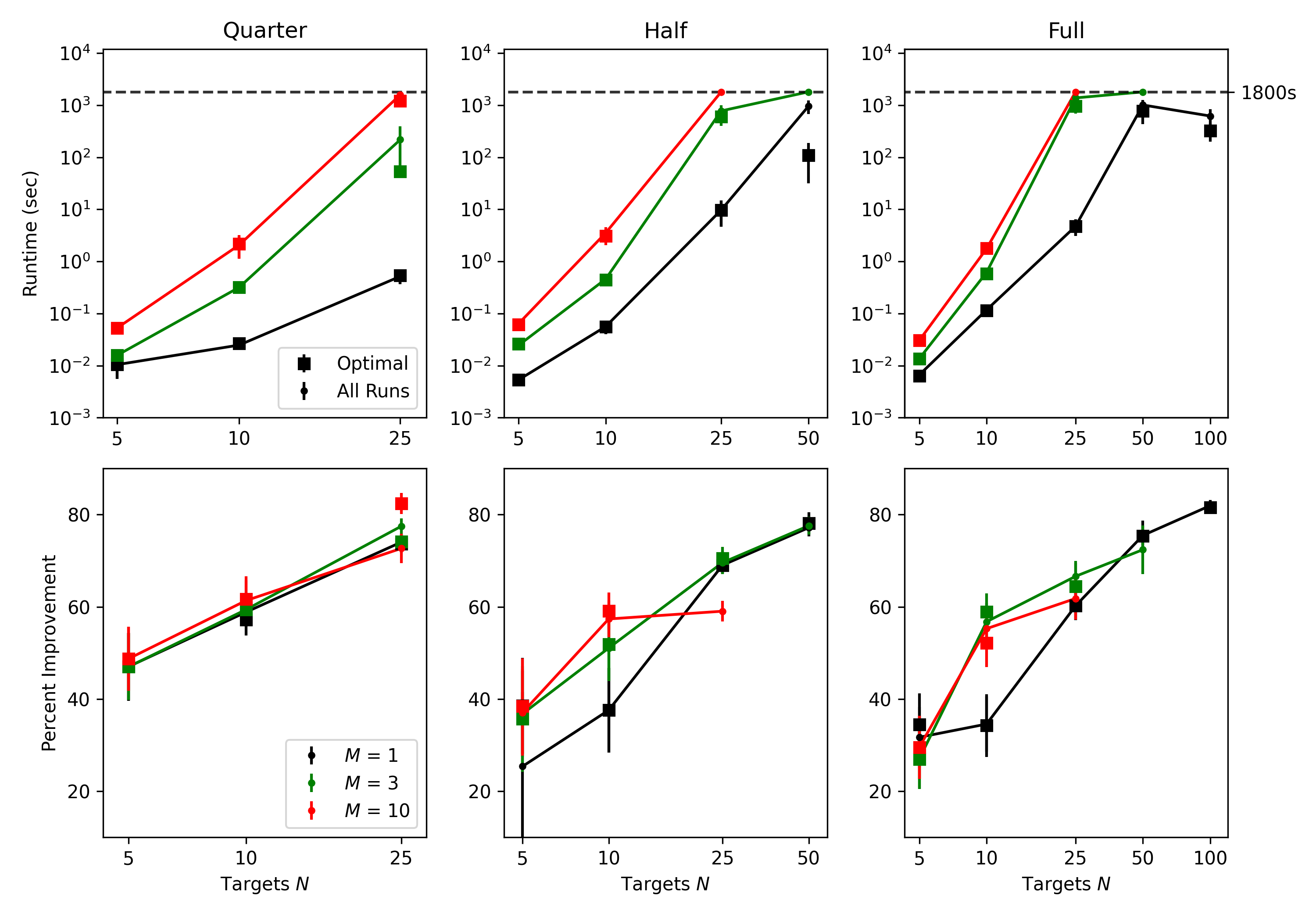}
    \caption{Top row: average value of $\Runtime$ for \TTPGLOBAL when scheduling each duration type $D$, with varying values of $M$. In the quarter night case, \TTPGLOBAL can schedule all 25 observations for $M<10$, but struggles to reach optimality at $M=10$. In the half night case, only the $M=1$ model achieves optimality for $N=50$. For the full night, \TTPGLOBAL handles $N=100$ far better than expected for the static case in comparison to $N=50$. Bottom row: average relative improvement $\RelRed_{\real}$ for each value of $D$ across the model types. For all $D$ and the respective maximum values of $N$, \TTPGLOBAL can reliably reduce slews by around a factor of 5. For small $N$, there is some benefit to using $M>1$. For larger models, the risk of long slews in the $M=1$ case has little impact on $\RelRed_{\real}$.}
    \label{fig:ttpglobal}
\end{figure*}

Table~\ref{table:ttpglobal_results} summarizes these statistics for the ten experiments conducted at each combination of  $N$, $M$ and $D$. We report the number of experiments where Gurobi found a feasible solution, $\NumFeas$, and a provably optimal solution $\NumOptimal$. For the feasible set, we report the average values of  $\SlewTime_{\tau}$, $\RelRed_{\tau}$, $\SlewTime_{\real}$, and $\RelRed_{\real}$. for each combination of $N$, $M$ and $D$, where the average is taken over the $\NumFeas$ instances for which a feasible solution was found. For example, for $(\text{Half}, 50, 3)$, $\NumFeas$ is 8, indicating that Gurobi found a feasible schedule in only 8 out of the 10 instances; consequently, the value of 12.7 for $\SlewTime_{\tau}$ is the average slew time over those 8 feasible schedules. We show the average $\Runtime$ and $\RelRed_{\real}$ values for different problem sizes in Figure~\ref{fig:ttpglobal}.

\begin{table*}
\begin{tabular}{lrrrrrrrrr}
\centering
$D$ & $N$ & $M$ & $\NumFeas$ & $\NumOptimal$ & $\Runtime$ & $\SlewTime_{\tau}$  & $\RelRed_{\tau}$ & $\SlewTime_{\real}$  & $\RelRed_{\real}$  \\ 
& & & & & (s) & (min) & (\%) & (min) & (\%) \\
\midrule
Quarter &  5 &  1 & 10 & 10 & 0.0 & 4.1 & 59.1 & 5.3 & 47.0 \\ 
Quarter &  5 &  3 & 10 & 10 & 0.0 & 3.7 & 63.3 & 5.3 & 47.0 \\ 
Quarter &  5 & 10 & 10 & 10 & 0.1 & 3.5 & 64.5 & 5.1 & 48.7 \\ 
Quarter & 10 &  1 & 10 & 10 & 0.0 & 5.8 & 70.9 & 8.2 & 58.8 \\ 
Quarter & 10 &  3 & 10 & 10 & 0.3 & 5.4 & 73.2 & 8.1 & 59.4 \\ 
Quarter & 10 & 10 & 10 & 10 & 2.1 & 5.3 & 73.7 & 7.7 & 61.4 \\ 
Quarter & 25 &  1 & 10 & 10 & 0.5 & 8.1 & 83.8 & 13.0 & 74.0 \\ 
Quarter & 25 &  3 & 10 &  9 & 217.7 & 7.8 & 84.4 & 11.3 & 77.5 \\ 
Quarter & 25 & 10 &  8 &  3 & 1625.2 & 9.6 & 80.8 & 13.7 & 72.7 \\ 
Half &  5 &  1 & 10 & 10 & 0.0 & 4.9 & 50.8 & 7.5 & 25.4 \\ 
Half &  5 &  3 & 10 & 10 & 0.0 & 4.7 & 53.2 & 6.3 & 36.8 \\ 
Half &  5 & 10 & 10 & 10 & 0.1 & 4.4 & 55.5 & 6.3 & 37.1 \\ 
Half & 10 &  1 & 10 & 10 & 0.1 & 6.5 & 67.3 & 12.5 & 37.6 \\ 
Half & 10 &  3 & 10 & 10 & 0.5 & 5.8 & 70.8 & 9.8 & 51.1 \\ 
Half & 10 & 10 & 10 & 10 & 3.5 & 5.5 & 72.4 & 8.5 & 57.4 \\ 
Half & 25 &  1 & 10 & 10 & 9.7 & 9.2 & 81.5 & 15.5 & 69.0 \\ 
Half & 25 &  3 & 10 &  8 & 776.8 & 8.5 & 83.0 & 15.2 & 69.6 \\ 
Half & 25 & 10 &  8 &  0 & 1800.0 & 12.7 & 74.6 & 20.5 & 59.1 \\ 
Half & 50 &  1 & 10 &  5 & 954.7 & 12.1 & 87.9 & 22.8 & 77.2 \\ 
Half & 50 &  3 &  8 &  0 & 1800.0 & 12.7 & 87.3 & 22.4 & 77.6 \\ 
Half & 50 & 10 &  0 &  0 & 1800.1 & -- & -- & -- & -- \\ 
Full &  5 &  1 &  9 &  9 & 0.0 & 6.0 & 39.7 & 6.8 & 31.7 \\ 
Full &  5 &  3 &  9 &  9 & 0.0 & 5.3 & 47.0 & 7.3 & 27.0 \\ 
Full &  5 & 10 &  9 &  9 & 0.0 & 5.3 & 47.3 & 7.0 & 29.5 \\ 
Full & 10 &  1 &  9 &  9 & 0.1 & 8.3 & 58.6 & 13.1 & 34.6 \\ 
Full & 10 &  3 &  9 &  9 & 0.6 & 7.4 & 62.8 & 8.6 & 56.8 \\ 
Full & 10 & 10 &  9 &  9 & 1.6 & 6.5 & 67.6 & 8.9 & 55.3 \\ 
Full & 25 &  1 & 10 & 10 & 4.7 & 10.4 & 79.1 & 19.9 & 60.2 \\ 
Full & 25 &  3 & 10 &  5 & 1379.4 & 9.6 & 80.7 & 16.7 & 66.6 \\ 
Full & 25 & 10 & 10 &  0 & 1800.0 & 12.3 & 75.3 & 19.1 & 61.8 \\ 
Full & 50 &  1 & 10 &  7 & 1009.6 & 13.2 & 86.8 & 24.5 & 75.5 \\ 
Full & 50 &  3 &  3 &  0 & 1800.0 & 18.8 & 81.2 & 27.6 & 72.4 \\ 
Full & 50 & 10 &  0 &  0 & 1800.0 & -- & -- & -- & -- \\ 
Full & 100 &  1 & 10 &  8 & 619.8 & 16.4 & 91.8 & 36.1 & 82.0 \\ 
Full & 100 &  3 &  0 &  0 & 1800.1 & --  & -- & -- & -- \\ 
Full & 100 & 10 &  0 &  0 & 1800.0 & -- & -- & -- & -- \\  \bottomrule
\end{tabular}
\caption{Computational results of \TTPGLOBAL for different values of $D$, $N$ and $M$. \emph{Note}: ``--'' indicates that the metric could not be calculated due to Gurobi not being able to obtain a feasible schedule for any of the ten instances. For $D = \text{Full}$, $N = 5$ and $D = \text{Full}$, $N = 10$, one instance was determined to be infeasible. The goal of our experiment is to devise target lists and exposure times that are feasible in the limit of large N, however for small N this is not strictly guaranteed.  \label{table:ttpglobal_results}}
\end{table*}

\section{Discussion} 
\label{sec:discussion}

We may draw a number of conclusions about the suitability of \TTPGLOBAL for the TTP from Table~\ref{table:ttpglobal_results} and Figure~\ref{fig:ttpglobal}. Gurobi generally found a feasible schedule when the number of targets $N \leq 25$. For $N \geq 50$ targets, our ability to find a feasible solution depended sensitively on the number of slots. Gurobi found a feasible schedule for all ten instances when $M = 1$, for some when $M = 3$, and for none when $M = 10$. 

Not surprisingly, runtime was a strong function of $N$ and $M$ as can be clearly seen in Figure~\ref{fig:ttpglobal}. For example, Gurobi found optimal solution for all $(D,N,M) = (\text{Full},25,1)$ experiments with an average $\Runtime$ of 4.7 seconds. In contrast, the $(\text{Full},25,10)$ experiments found no optimal solutions in the 1800 second time limit. For the largest $(\text{Full},100,3)$ experiments, not even a feasible solution was found in the time limit. We find that the $M=1$ case scales well into the realm of $N=100$, solving even faster than the $N=50$ case. While this goes against our initial intuition, we suspect this behavior is the result of parameter tuning by Gurobi to accommodate larger models.

When we do find a feasible schedule, it is significantly more efficient relative to the $2N$ baseline. For example, for the $(D,N,M) = (\text{Half},50,3)$ experiments, the average $\SlewTime_{\real}$ is 22.4 minutes compared to the $2N = 100$ minute benchmark, a reduction of $\RelRed_{\real}$ = 77.6\%. For the $(\text{Full},100,1)$ set of instances, average $\SlewTime_{\real}$ is 36.1 minutes, which is a reduction of 82.0\% relative to the $2N = 200$ minute benchmark. We note that in all cases, $\SlewTime_{\tau}$ is less than $\SlewTime_{\real}$. For example, for the same $D = \text{Full}$, $N = 100$, $M = 1$ set of instances, $\SlewTime_{\tau}$ is a mere 16.4 minutes, which is a reduction of $\RelRed_{\tau}$ = 91.8\% relative to the 200 minute worst-case value. The difference between $\RelRed_{\real}$ and $\RelRed_{\tau}$ stems from our piece-wise slew approximation, i.e. the choice of $M$ for each model.

In the quarter night models, we find little improvement from using higher values of $M$, since the slews vary minimally with respect to time. For the longer durations, we see some benefit for higher $M$ in the simpler $N<25$ cases. For $N\ge25$, the higher $M$ models become too complex to be solved to optimality in the time limit (leaving better slews in question), but the $M=1$ models continue to demonstrate extremely efficient slews despite the higher expected variability.

In most cases, increasing the value of $M$ did not demonstrate a significant advantage over the static models. For the high $N$ models, the optimizer was often not able to construct a feasible solution for $M>1$. For scheduling full nights of observations, the $M=1$ case demonstrates dramatic improvement in slew times by up to a factor of 5, and increasing $M$ provides more computational complexity than can be handled by our global algorithm in the time limit. For most cases, we would recommend $M=1$ be the standard due to its exceptionally fast runtime.

In our numerical experiments we attempted to solve \TTPGLOBAL to a provably optimal solution by branch-and-bound. For large numbers of targets or finely discretized slew tensors this approach may not be computationally tractable. We note that many heuristic solutions to the TSP have been developed that achieve near-optimal solutions. We suspect that analogous high-quality heuristic solutions exist for the TTP, which may be equipped to solve the $M>1$ models for much higher $N$. We develop one such procedure in the Appendix for comparison with our global solution.

\section{Conclusions} 
\label{sec:conclusion}

In this work, we addressed the challenge of determining the most efficient ordering of a set of exposures at a telescope. We formulated the Traveling Telescope Problem (TTP) as a mixed-integer linear program \TTPGLOBAL and solved it using a standard commercial optimizer for problem sizes as large as 100 targets in $\sim 10$ minutes using modest computational resources. The speed of \TTPGLOBAL means it can be run dynamically throughout the night and respond to real-time changes in target accessibility from weather. Further work is needed to develop algorithms that can solve the TTP for substantially larger sets of targets or substantially finer time-resolution in slew overheads. Local searches initialized with an heuristic solution may prove fruitful. We hope that algorithms like the one described here can assist or automate scheduling, save human effort and increase the scientific productivity of astronomical surveys.

\begin{acknowledgments}
L.H. \& E.A.P. acknowledge support from the following sources: Heising-Simons Foundation Grant \#2022-3832. V.V.M. acknowledges support from the UCLA Anderson School of Management. We are grateful for enlightening conversations with Eric Bellm. 

% Acknowledge conversations with...

\end{acknowledgments}



\software{astropy \citep{2013A&A...558A..33A,2018AJ....156..123A}, numpy \citep{numpy}, pandas \citep{pandas}, gurobi \citep{gurobi}, matplotlib \citep{matplotlib}}


\appendix
\section{Variables}
\label{sec:variables}
Table~\ref{table:variables} lists the variables used in all preceding sections of this paper, along with their first usage:

\begin{table}[h!]
    \centering
    \caption{Symbols Used}\label{table:variables}
    \begin{tabular}{|c|l|c|}
    \hline
    \textbf{Symbol} & \textbf{Definition} & \textbf{Section} \\
    \hline
    $C$ & Small normalization constant used in objective function. Ensures the scheduling of additional & \\
    & observations is prioritized & \ref{sec:nightobj}\\
    $i,j,k$ & Indices for arbitrary nodes & \ref{sec:slots} \\
    $m$ & Index for an arbitrary time slot & \ref{sec:slots}\\
    $M$ & The number of time slots where slews are assumed constant. & \ref{sec:slots}\\
    $N$ & The number of celestial targets assigned to the TTP & \ref{sec:slots} \\
    $t_{i}^\mathrm{e}$ & The earliest time value at which the node $i$ can be departed based on observability constraints & \ref{sec:slots}\\
    $t_{i}^\mathrm{\ell}$ & The latest time value at which the node $i$ can be departed based on observability constraints & \ref{sec:slots}\\
    $t_{i}$ & A continuous variable indicating the time value of the departure from node $i$ & \ref{sec:nightvar} \\
    $w_{m}$ & The time value at which the slot $m$ begins & \ref{sec:slots}\\
    $X_{ijm}$ & A binary decision variable indicating the flow state between nodes $i,j$ during the time slot $m$. & \\
    & Holds 1 if a slew takes place, and 0 otherwise & \ref{sec:nightvar}\\
    $Y_{i}$ & A binary decision variable that indicates the visitation of the node $i$. Holds 1 if the node is & \\
    & visited, and 0 otherwise & \ref{sec:nightvar}\\
    $\tau_{i}^\mathrm{exp}$ & The exposure duration of the target node $i$ & \ref{sec:slots} \\
    $\tau_{ijm}^\mathrm{slew}$ & Travel time between the nodes $i,j$ during the time slot $m$ & \ref{sec:slots}\\
    \hline
    \end{tabular}
\end{table}

\section{Variants of the Travelling Telescope Problem}
\label{sec:variants}

\subsection{Consecutive Targeting}
One may require two exposures $i$ and $i'$ be scheduled back-to-back, such as a science observation and a calibration observation. Such linked observations may be specified via two additional constraints:

\begin{align}
    Y_{i} + Y_{i'} = \,& 2\left(\sum_{m=0}^{M-1} X_{ii'm} + \sum_{m=0}^{M-1} X_{i'im}\right) \\
    Y_{i} = \,& Y_{i'}.
\end{align}
The first constraint ensures that if both observations take place, the directed tour must pass directly from $i$ to $i'$ or vice versa. The second constraint ensures both observations or neither observation occur.

\subsection{Intra-Night Spacing Requirements}
\label{sec:cadence}
One may wish to enforce a minimum interval between two exposures. A common example occurs in time-series monitoring where multiple observations of the same target occur during the same night, subject to a minimum separation. We accomplish this by letting $N$ correspond to the total number of \textit{exposures} to be collected across all targets. For simplicity let us assign exposure indices such that exposures of the same target occur consecutively in the total exposure list $\{ 1,\ldots,N \}$. That is, if the target requires $n^\mathrm{exp}$ individual exposures, the node indices $\{ \kappa,\kappa +1,\ldots,\kappa+n^\mathrm{exp} -1\}$ correspond to that target for some value $\kappa$.

With the repeat observations specified, we enforce a minimum interval via the following  constraint:
\begin{equation}
\begin{aligned}
\sum_{j = 1}^{N+1} \sum_{m = 0}^{M-1} t_{ijm} \ge \sum_{j = 1}^{N+1} \sum_{m = 0}^{M-1} t_{i-1,j,m} + Y_{i} \tau^\mathrm{sep}, \quad \forall\ i = \kappa+1,\ldots,\kappa+n^\mathrm{exp} -1
\end{aligned}
\end{equation}
Subsequent exposures of a given target must not end until at least $\tau^\mathrm{sep}$ has passed since the previous exposure ended. For example, if the linked exposures of a target have index $i$ = 5 and 6, exposure 6 may not end until at least $\tau^\mathrm{sep}$ has passed since exposure 5 ended.

%\newpage
\clearpage

\section{Local Search Heuristic}
Here, we develop an alternative formulation to the TTP which uses a local search heuristic. We refer to this as \textsf{TTP-LS} to distinguish it from \TTPGLOBAL which searches the global solution space.

\subsection{Algorithm Description}
\label{subsec:localsearch_description}

In the TTP, there are two sets of decisions that need to be made simultaneously. One is the sequence in which the targets will be visited. For example, with $N = 5$, we have to choose between $5 \to 1 \to 4 \to 2 \to 3$, $1 \to 3 \to 2 \to 4 \to 5$, and so on. The other set of decisions involves the timing slews. This involves deciding a (continuous) time $t_i$ within the observing duration D to slew and the corresponding slot $m$. Even with a fixed sequence exposures, this is a non-trival task. As a result, making both sets of decisions under the umbrella of a single MILP formulation is computationally demanding. 

This suggests an alternate approach to the TTP that decouples the sequence decision from the timing decision. Suppose that the sequence of targets is fixed. When should the telescope slew in order to minimize slew times? Let $\sigma$ denote the sequence of targets, which is a bijective function $\sigma: \{1,\dots,N\} \to \{1,\dots, N\}$, and let the minimum total slew time be denoted by the function $F$, so that $F(\sigma)$ is the minimum total slew time that one would obtain from following the sequence $\sigma$. Let $\Sigma$ denote the set of all such sequences. For example, for $N = 5$ targets and the sequence $5 \to 1 \to 4 \to 2 \to 3$, the corresponding $\sigma$ is 
\begin{align*}
\sigma(1) & = 5 \\
\sigma(2) & = 1 \\
\sigma(3) & = 4 \\
\sigma(4) & = 2 \\ 
\sigma(5) & = 3
\end{align*}
The TTP can then be abstractly formulated as 
\begin{equation*}
\min_{\sigma \in \Sigma} F(\sigma),
\end{equation*}
which is an optimization problem over sequences in $\Sigma$. As written, this is not a problem that can be readily provided to any commercial solver, but because of the discrete nature of $\Sigma$, it can potentially be solved using local search. Let $z \in \{1,\dots, N\}$ and $z' \in \{1,\dots,z-1,z+1,\dots,N\}$ be positions in the sequence, and let $\sigma^{z \leftrightarrow z'}$ denote the sequence obtained by swapping the targets in positions $z$ and $z'$; that is, $\sigma^{z \leftrightarrow z'}$ is the unique sequence such that 
\begin{align*}
\sigma(z) & = \sigma^{z \leftrightarrow z'}(z'), \\
\sigma(z') & = \sigma^{z \leftrightarrow z'}(z), \\
\sigma(z'') & = \sigma^{z \leftrightarrow z'}(z'') \quad \forall z'' \in \{1,\dots, N\} \setminus \{z, z' \}. 
\end{align*}

Let $\Ncal_z(\sigma)$ denote the set of neighboring sequences of $\sigma$ obtained by swapping the target at position $z$ with a target at any other position:
\begin{equation*}
\Ncal_z(\sigma) = \left\{ \sigma' \in \Sigma \ \vline \ \begin{array}{l} \sigma' = \sigma^{z \leftrightarrow z'} \\ \text{for some} \ z' \in \{1,\dots,z-1,z+1,\dots,N\} \end{array} \right\}
\end{equation*}
With this definition, our local search algorithm can be formally described as Algorithm~\ref{alg:local_search}. 

\begin{algorithm}
\begin{algorithmic}[1]
\REQUIRE Initial sequence $\sigma \in \Sigma$. 
\STATE Set $\Ucal \gets \{1,\dots, N\}$. 
\WHILE{$|\Ucal| > 0$}
		\STATE Select $z \in \Ucal$; set $\Ucal \gets \Ucal \setminus \{z \}$
		\STATE Set $\sigma^* \gets \arg \min_{\sigma' \in \Ncal_z(\sigma)} F(\sigma')$
        \STATE Set $F' \gets \min_{\sigma' \in \Ncal_z(\sigma)} F(\sigma')$
		\IF{$F' < F(\sigma)$}
			\STATE Set $\sigma \gets \sigma^*$,
			\STATE Set $\Ucal \gets \{1,\dots, z-1,z+1, \dots, N\}$
		\ENDIF
\ENDWHILE
\RETURN Locally optimal sequence $\sigma$
\end{algorithmic}
\caption{Pseudocode of local search procedure. \label{alg:local_search}}
\end{algorithm}

In words, we begin from some initial sequence $\sigma$. We use $\Ucal$ to denote the set of sequence positions which have not yet tried to modify. As long as there is at least one sequence position we have not tried to change, we pick a sequence position $z$, and calculate the best neighboring sequence $\sigma^*$ obtained by swapping the target at position $z$ with the target at any other position. If the objective value $F'$ of the best neighboring sequence improves on the current objective value $F(\sigma)$, we replace $\sigma$ with $\sigma^*$, and we reset $\Ucal$ to be the set of all positions. Otherwise, if we do not make improvement in an iteration of the while loop, then $\Ucal$ will be reduced by one member. If $N$ such iterations occur, then $\Ucal$ will be empty, and we will have ascertained that there is no neighboring sequence we can move to in order to reduce the objective value; in other words, $\sigma$ is a locally optimal sequence. We note that this heuristic is similar to the 2-OPT heuristic \citep{croes1958method} for the classical TSP problem, which involves eliminating two edges in a TSP tour and reconnecting the tour so that the edges are swapped. The main difference comes from the function $F$, which calculates the minimum total slew time when one assigns the targets in the sequence to slots optimally.

Before this algorithm can be deployed, we must specify how to compute $F(\sigma)$. The function value $F(\sigma)$ for a fixed sequence $\sigma$ can be calculated by solving a MILP. While this MILP shares some similarities with the TTP problem described in Section~\ref{sec:ttp}, it is a simpler because target sequence is fixed and  ``baked in'' to the optimization problem. This smaller MILP is a subroutine in Algorithm~\ref{alg:local_search}. We provide further details on this integer program in Section~\ref{subsec:localsearch_F}. 

We must also consider sequences of targets that are infeasible. In some cases, for a fixed sequence $\sigma$ of targets, it may not be possible to make the timing decisions and the slot decisions in way that respects the accessibility windows and slot bounds. In such cases, the MILP that defines the function $F(\cdot)$ will be infeasible. We can extend the definition of $F(\cdot)$ so that $F(\sigma) = +\infty$ if the corresponding MILP is infeasible for sequence $\sigma$; since Algorithm~\ref{alg:local_search} is always choosing the neighboring sequence with the lowest value of $F$, this ensures that Algorithm~\ref{alg:local_search} will never replace the current sequence with one that is infeasible. 

However, even with this fix, one problem that still remains is if the initial sequence $\sigma$ and all neighboring sequences of that initial sequence are infeasible. In this case, Algorithm~\ref{alg:local_search} will not return a feasible sequence, as it will simply terminate with the current sequence. This is a serious issue, because it is not straightforward to identify a sequence of targets for which the TTP problem constraints can be perfectly satisfied. In Section~\ref{subsec:localsearch_feasibility}, we present a feasibility heuristic for identifying such a sequence.

\subsection{Calculating Minimum Total Slew Time for a Fixed Sequence of Targets}
\label{subsec:localsearch_F}

As noted in the previous section, a key component of Algorithm~\ref{alg:local_search} is the function $F$, which maps a sequence $\sigma$ to a minimum slew time $F(\sigma)$. We use $z$ to denote the index of a position in this sequence $\sigma$. For targets, $z$ will range in $\{1,2,\dots, N\}$. With a slight abuse of notation, we will use $z = 0$ to denote the start node of the telescope, and assume that $\sigma(0) = 0$; similarly $\sigma(N+1) = N+1$ at the final node. Thus, $z$ can take any value in $\{0,1,\dots,N+1\}$. 

Let $Y^{\at}_{z,m}$ be a binary decision variable that is 1 if the telescope departs the target at position $z \in \{0,1,\dots,N+1\}$ in the sequence in slot $m$, and 0 otherwise. Let $Y^{\by}_{z,m}$ be a binary decision variable that is 1 if the telescope reaches slot $m$ by position $z \in \{0,1,\dots,N+1\}$ in the sequence and 0 otherwise. Let $t_z$ denote the departure time of the telescope from the node at position $z$. The function $F$ is obtained by solving the following MILP:
{\allowdisplaybreaks
\begin{subequations}
\begin{alignat}{2}
& \text{minimize} & \quad &  \sum_{z = 1}^N \sum_{m=0}^{M-1}  \tau^{\slew}_{ \sigma(z), \sigma(z+1), m} Y^{\at}_{z,m} \\
& \text{subject to} & & Y^{\by}_{0,0} = 1, \\
& & & Y^{\by}_{z,m} \leq Y^{\by}_{z+1, m}, \quad \forall \ z = 0,1,\dots, N , \ m = 0,1,\dots,M-1, \label{prob:seq_by_successive_z}\\
& & & Y^{\by}_{z,m+1} \leq Y^{\by}_{z,m}, \quad \forall \ z = 0,1,\dots,N+1, \ m = 0,1,\dots,M-2, \label{prob:seq_by_successive_m}\\ 
& & & Y^{\at}_{z,m} = Y^{\by}_{z,m} - Y^{\by}_{z,m+1}, \quad \forall \ z = 0,1,\dots,N+1, \ m = 0,1,\dots,M-2, \label{prob:seq_by_at_1} \\
& & & Y^{\at}_{z,M-1} = Y^{\by}_{z,M-1}, \quad \forall \ z = 0,1,\dots,N+1, \label{prob:seq_by_at_2} \\
& & & t_z \geq t_{z-1} + \sum_{m=0}^{M-1} \tau^{\slew}_{\sigma(z-1), \sigma(z), m} \cdot Y^{\at}_{z-1,m} + \tau^\mathrm{exp}_{\sigma(z)}, \quad \forall z = 1,2,\dots,N+1, \label{prob:seq_time_dynamics}\\
& & & t_z \geq \sum_{m=0}^{M-1} w_m \cdot Y^{\at}_{z,m}, \quad \forall z = 0,1,\dots,N+1, \label{prob:seq_slot_upper} \\
& & & t_z \leq \sum_{m=0}^{M-1} w_{m+1} \cdot Y^{\at}_{z,m}, \quad \forall z = 0,1,\dots,N+1, \label{prob:seq_slot_lower} \\
& & & t_z \geq t^e_{\sigma(z)}, \quad \forall \ z = 0,1,\dots,N+1,  \label{prob:seq_rise}\\
& & & t_z \leq t^\ell_{\sigma(z)}, \quad \forall \ z = 0,1,\dots,N+1, \label{prob:seq_set} \\
& & & Y^{\by}_{z,m} \in \{0,1\}, \quad \forall \ z = 0,1,\dots,N+1, \ m = 0,1,\dots,M-1, \label{prob:seq_q_by_binary} \\
& & & Y^{\at}_{z,m} \in \{0,1\}, \quad \forall \ z = 0,1,\dots,N+1, \ m = 0,1,\dots,M-1. \label{prob:seq_q_at_binary}
\end{alignat}%
\label{prob:seq}
\end{subequations}
}
In order of appearance, the constraints have the following meaning. Constraint~\eqref{prob:seq_by_successive_z} requires that if we have reached slot $m$ by position $z$, then it must be the case that we have reached slot $m$ by position $z+1$. Constraint~\eqref{prob:seq_by_successive_m} requires that if we have reached slot $m+1$ by position $z$, then we must have reached slot $m$ by position $z$. Constraints~\eqref{prob:seq_by_at_1} and \eqref{prob:seq_by_at_2} link the $Y^{\by}$ and $Y^{\at}$ variables; constraint~\eqref{prob:seq_by_at_1} means that we are in slot $m$ at position $z$ if and only if we have reached slot $m$ by position $z$ ($Y^{\by}_{z,m} = 1$) and have not reached slot $m+1$ by position $z$ ($Y^{\by}_{z,m+1} = 0$). Constraint~\eqref{prob:seq_by_at_2} similarly requires that we are in slot $M-1$ at position $z$ if and only if we have reached slot $M-1$ by position $z$. Constraint~\eqref{prob:seq_time_dynamics} requires that the departure time from the target in position $z$ is at least the slew time that is realized departing from the target in slot $z-1$ plus the exposure time of the target in position $z$. Constraints~\eqref{prob:seq_slot_upper} and \eqref{prob:seq_slot_lower} ensure that the departure time of each position $z$ is within the lower and upper bounds of that position's assigned slot, while constraints~\eqref{prob:seq_rise} and \eqref{prob:seq_set} ensure that the departure time from each position $z \in \{1,\dots,N\}$ is within the rise and set times for the target in that position ($t^e_{\sigma(z)}$ and $t^\ell_{\sigma(z)}$ respectively). The last two constraints enforce that the $Y^{\at}$ and $Y^{\by}$ variables are binary. 

The MILP problem~\eqref{prob:seq} is essentially the TTP problem of Section~\ref{sec:ttp}, restricted to a particular sequence $\sigma$. Essentially, this formulation decides when each target's departure time will be and to what slots the different positions in the sequence will be allocated. Importantly, the sequence of targets is not a decision variable as it is in the original TTP model; it is a fixed input that is provided by the user. 

Many of the constraints are direct analogs of constraints that appear in the \TTPGLOBAL formulation. For example, \eqref{prob:seq_rise} and \eqref{prob:seq_set} model the rise and set time constraints for each target in each position, mirroring constraint~\eqref{eq:windows} of the TTP MILP. As another example, \eqref{prob:seq_slot_lower} and \eqref{prob:seq_slot_upper} model the lower and upper bounds of each slot, similarly to constraint~\eqref{eq:tmin}. Note that because the sequence of targets is fixed, many of the constraints from the full TTP MILP can be simplified, and other decisions, such as the departure times and in which slot each target is being departed from, can be expressed more efficiently using different decision variables. Specifically, the decision of which slot each position is assigned to is captured by the $Y^{\by}_{z,m}$ decision variables. Here, we remark that these variables are an example of the incremental encoding technique in integer programming, which enhances the efficiency of branching in the branch-and-bound algorithm that is the cornerstone of integer programming solvers. We refer interested readers to the review paper of \cite{vielma2015mixed}, and to \cite{bertsimas2011integer}, \cite{bertsimas2019airlift} and \cite{mivsic2020optimization} for examples of applications of this technique in air traffic control, vehicle routing and optimization over trained machine learning models. 

As a result, problem~\eqref{prob:seq} is much easier to solve than the full TTP MILP. We found that that Gurobi could determine the exact optimal solution to this problem in under a second with a single thread.

\subsection{Feasibility Algorithm}
\label{subsec:localsearch_feasibility}

The local search approach described above may fail if it initialized at an infeasible sequence whose neighbors are also all infeasible. In order for Algorithm~\ref{alg:local_search} to return a sequence that can be implemented, the initial sequence must be one for which the minimum slew problem~\eqref{prob:seq} is feasible. Given the combinatorical complexity of the TTP, it is unlikely that one would randomly select a target sequence that would result in problem~\eqref{prob:seq} being feasible. 

Thus motivated, we present in this section an algorithm that, starting from any sequence, seeks to return a feasible sequence. We note that this algorithm is a heuristic, and is not guaranteed to succeed. Nevertheless, our numerical results in Section~\ref{subsec:results_LS} indicate that this heuristic is generally very effective. 

At a very high level, the algorithm we will propose resembles our local search procedure, Algorithm~\ref{alg:local_search}, in that it starts from a sequence $\sigma$ and makes moves to neighboring sequences. The key difference is the objective function that is used. Instead of using the function $F$, the feasibility algorithm first seeks to locally optimize a function $G_1$, followed by a function $G_2$. The function $G_1(\sigma)$ measures, for the sequence $\sigma$, the smallest violation of the slot window constraints~\eqref{prob:seq_slot_upper} and \eqref{prob:seq_slot_lower} that can be attained when we choose the departure times and the slots to which each target is assigned to. This violation is a nonnegative quantity; a positive value implies that we are unable to satisfy all of the constraints, i.e., at least one constraint in constraint sets~\eqref{prob:seq_slot_upper} and \eqref{prob:seq_slot_lower} is violated. A value of zero implies that all of the constraints in the two constraint sets are satisfied. The function $G_2(\sigma)$ similarly measures the smallest possible violation of the visibility  constraints~\eqref{prob:seq_rise} and \eqref{prob:seq_set} when we choose the departure times and the slots. Again, a positive value implies that at least one constraint in the constraint sets \eqref{prob:seq_rise} and \eqref{prob:seq_set} is violated, while a value of zero implies we can satisfy all of the constraints defined by \eqref{prob:seq_rise} and \eqref{prob:seq_set}.

$G_1$ is defined by the following MILP:
{\allowdisplaybreaks
\begin{subequations}
\begin{alignat}{2}
& \text{minimize} & \quad &  \sum_{z = 0}^{N+1} \epsilon^{\slote}_{z} + \sum_{z = 0}^{N+1} \epsilon^{\slotell}_{z} \\
& \text{subject to} & & Y^{\by}_{0,0} = 1, \label{prob:G1_by_0_0}\\
& & & Y^{\by}_{z,m} \leq Y^{\by}_{z+1, m}, \quad \forall \ z = 0,1,\dots, N , \ m = 0,1,\dots,M-1, \label{prob:G1_by_successive_z}\\
& & & Y^{\by}_{z,m+1} \leq Y^{\by}_{z,m}, \quad \forall \ z = 0,1,\dots,N+1, \ m = 0,1,\dots,M-2, \label{prob:G1_by_successive_m}\\ \\
& & & Y^{\at}_{z,m} = Y^{\by}_{z,m} - Y^{\by}_{z,m+1}, \quad \forall \ z = 0,1,\dots,N+1, \ m = 0,1,\dots,M-2, \label{prob:G1_by_at_1} \\
& & & Y^{\at}_{z,M-1} = Y^{\by}_{z,M-1}, \quad \forall \ z = 0,1,\dots,N+1, \label{prob:G1_seq_by_at_2} \\
& & & t_z \geq t_{z-1} + \sum_{m=0}^{M-1} \tau^{\slew}_{\sigma(z-1), \sigma(z), m} \cdot Y^{\at}_{z-1,m} + \tau^\mathrm{exp}_{\sigma(z)}, \quad \forall z = 1,2,\dots,N+1, \label{prob:G1_seq_time_dynamics}\\
& & & t_z \geq \sum_{m=0}^{M-1} w_{m} \cdot Y^{\at}_{z,m} - \epsilon^{\slote}_z, \quad \forall z = 0,1,\dots,N+1, \label{prob:G1_slot_lower} \\
& & & t_z \leq \sum_{m=0}^{M-1} w_{m+1} \cdot Y^{\at}_{z,m} + \epsilon^{\slotell}_{z}, \quad \forall z = 0,1,\dots,N+1, \label{prob:G1_slot_upper} \\
& & & t_z \geq t^e_{\sigma(z)} - \epsilon^\mathrm{e}_z , \quad \forall \ z = 0,1,\dots,N+1,  \label{prob:G1_rise}\\
& & & t_z \leq t^\ell_{\sigma(z)} + \epsilon^{\ell}_z, \quad \forall \ z = 0,1,\dots,N+1, \label{prob:G1_set} \\
& & & Y^{\by}_{z,m} \in \{0,1\}, \quad \forall \ z = 0,1,\dots,N+1, \ m = 0,1,\dots,M-1, \label{prob:G1_q_by_binary} \\
& & & Y^{\at}_{z,m} \in \{0,1\}, \quad \forall \ z = 0,1,\dots,N+1, \ m = 0,1,\dots,M-1, \label{prob:G1_q_at_binary} \\
& & & \epsilon^\mathrm{e}_z, \epsilon^{\ell}_z, \epsilon^{\slotell}_z, \epsilon^{\slote}_z \geq 0, \quad \forall z = 0,1,\dots,N+1. \label{prob:G1_epsilon_def}
\end{alignat}
\label{prob:G1}%
\end{subequations}
}
Observe that this integer program is similar to the minimum slew problem~\eqref{prob:seq}, except that we now allow for violations of the constraints~\eqref{prob:seq_slot_upper}, \eqref{prob:seq_slot_lower}, \eqref{prob:seq_rise} and \eqref{prob:seq_set}. The main modifications are as follows. First, observe that in addition to the decision variables of problem~\eqref{prob:seq}, problem~\eqref{prob:G1} includes the decision variables $\epsilon^\mathrm{e}_z, \epsilon^{\ell}_z, \epsilon^{\slotell}_z, \epsilon^{\slote}_z$, which measure how much the rise, set, slot upper bound and slot lower bound constraints can be violated in time. 

Second, observe that constraints~\eqref{prob:G1_slot_lower} - \eqref{prob:G1_set} resemble constraints~\eqref{prob:seq_slot_upper} - \eqref{prob:seq_set}, except that the new $\epsilon^\mathrm{e}_z, \epsilon^{\ell}_z, \epsilon^{\slotell}_z, \epsilon^{\slote}_z$ appear. These new decision variables are bounded from below by zero and unbounded from above, so they effectively allow the optimizer to choose to not satisfy these constraints. For example, in the constraint $t_z \leq t^\ell_{\sigma(z)} + \epsilon^{\ell}_z$, for whatever value of $t_z$ we choose, we can always make the constraint satisfied by setting $\epsilon^{\ell}_z$ to be equal to any value greater than $\max\{0, t_z - t^\ell_{\sigma(z)} \}$; as a concrete example, if $t_z = 120$ and $t^\ell_{\sigma(z)} = 80$, then any $\epsilon^{\ell}_z \geq \max\{120 - 80, 0\} = 40$ will satisfy the constraint. 

Lastly, observe that the objective function is equal to the sum of the $\epsilon^{\slotell}_z$ and $\epsilon^{\slote}_z$ variables. Thus, the optimizer seeks to find the assignments of sequence positions to slots and the departure times so as to minimize how much the slot lower and upper bound constraints from problem~\eqref{prob:seq} are violated. Observe that if the optimal objective value of problem~\eqref{prob:G1} is zero, then we have found a partially feasible solution to problem~\eqref{prob:seq} that satisfies all of the constraints, and in particular constraints~\eqref{prob:seq_slot_upper} and \eqref{prob:seq_slot_lower}, with the possible exception of the rise and set time constraints~\eqref{prob:seq_rise} and \eqref{prob:seq_set}. Importantly, note that no matter what sequence $\sigma$ one chooses, problem~\eqref{prob:G1} is always feasible. %We can fix the variables $Y^{\at}_{z,0} = 1$ for all $z$, $Y^{\at}_{z,1}

We now define the function $G_2$. The function $G_2$ is defined as the objective value of the following integer program, which is 
\begin{subequations}
\begin{alignat}{2}
& \text{minimize} & \quad &  \sum_{z = 0}^{N+1} \epsilon^\mathrm{e}_{z} + \sum_{z = 0}^{N+1} \epsilon^{\ell}_{z} \\
& \text{subject to} & & \text{constraints~\eqref{prob:G1_by_0_0} - \eqref{prob:G1_epsilon_def}}, \\
& & & \epsilon^{\slote}_z = 0, \quad \forall z = \{0,1,\dots,N+1\}, \\
& & & \epsilon^{\slotell}_z = 0, \quad \forall z = \{0,1,\dots,N+1\}.
\end{alignat}
\label{prob:G2}
\end{subequations}
Problem~\eqref{prob:G2} has the same structure as problem~\eqref{prob:G1}, except that we force the violation variables $\epsilon^{\slote}_z$ and $\epsilon^{\slotell}_z$ to zero; thus, we no longer allow for violations of the slot bound constraints~\eqref{prob:seq_slot_upper} and \eqref{prob:seq_slot_lower}. We do still allow for violations of the rise and set time constraints~\eqref{prob:seq_rise} and \eqref{prob:seq_set}. The objective function measures how much the rise and set time constraints are violated. Observe that if the objective value of problem~\eqref{prob:G2} is zero, then we have exactly verified that the minimum slew time MILP~\eqref{prob:seq} is feasible.

With problems~\eqref{prob:G1} and \eqref{prob:G2} defined, we can now define the feasibility algorithm, which we provide in Algorithm~\ref{alg:feasibility}. This algorithm works by first performing local search using the function $G_1$. If the local optimum is such that the value of $G_1$ is positive, then the algorithm terminates and returns that the problem is infeasible. Otherwise, if the value of $G_1$ is zero, then we continue to the next phase, in which we perform local search using the function $G_2$. If the local optimum of $G_2$ is positive, then the algorithm again terminates and returns that the problem is infeasible. Otherwise, if the value of $G_2$ is zero, then we have identified a feasible sequence. Note that Algorithm~\ref{alg:feasibility} is a heuristic and does not provably verify that problem~\eqref{prob:seq} is infeasible. If it returns ``Problem is infeasible'', it may not be the case that the minimum slew problem is actually infeasible.

\begin{algorithm}
\begin{algorithmic}[1]
\REQUIRE Initial sequence $\sigma \in \Sigma$. 
\STATE  \COMMENT{Phase 1: Minimization of $G_1$ (violation of slot lower and upper bound constraints)}
\STATE Set $\Ucal \gets \{1,\dots, N\}$. 
\WHILE{$|\Ucal| > 0$}
		\STATE Select $z \in \Ucal$; set $\Ucal \gets \Ucal \setminus \{z \}$.
		\STATE Calculate $\sigma^* \gets \arg \min_{\sigma' \in \Ncal_z(\sigma)} G_1(\sigma')$.
        \STATE Calculate $G_1' \gets \min_{\sigma' \in \Ncal_z(\sigma)} G_1(\sigma')$.
		\IF{$G_1' < G_1(\sigma)$}
			\STATE Set $\sigma \gets \sigma^*$.
			\STATE Set $\Ucal \gets \{1,\dots, z-1,z+1, \dots, N\}$.
		\ENDIF
\ENDWHILE
\IF{ $G_1(\sigma) > 0$ }
	\RETURN Problem is infeasible.
\ELSE
	\STATE \COMMENT{Phase 2: Minimization of $G_2$ (violation of rise and set time constraints)}
	\STATE Set $\Ucal \gets \{1,\dots, N\}$. 
	\WHILE{$|\Ucal| > 0$}
			\STATE Select $z \in \Ucal$; set $\Ucal \gets \Ucal \setminus \{z \}$.
			\STATE Calculate $\sigma^* \gets \arg \min_{\sigma' \in \Ncal_z(\sigma)} G_2(\sigma')$.
            \STATE Calculate $G_2' \gets \min_{\sigma' \in \Ncal_z(\sigma)} G_2(\sigma')$.
			\IF{$G_2' < G_2(\sigma)$}
				\STATE Set $\sigma \gets \sigma^*$.
				\STATE Set $\Ucal \gets \{1,\dots, z-1,z+1, \dots, N\}$.
			\ENDIF
	\ENDWHILE
	\IF{ $G_2(\sigma) > 0$ }
		\RETURN Problem is infeasible.
	\ELSE 
		\RETURN Feasible sequence $\sigma$.
	\ENDIF
\ENDIF
\end{algorithmic}
\caption{Pseudocode of feasibility procedure. \label{alg:feasibility}}
\end{algorithm}
\subsection{Computational results for local search heuristic} 
\label{subsec:results_LS}

We now present our results on our heuristic approach described above. We tested our approach on the same collection of 360 experiments described in Section~\ref{subsec:results_TTPGLOBAL}, and compute the same result metrics. We tested two variants of our local search procedure:
\begin{itemize}
\item \LSONE: Here, we execute our overall algorithm from a single random starting point, which we obtain by drawing a sequence $\sigma$ uniformly at random from all possible $N!$ sequences. 
\item \LSTEN: In the second variant, we execute our overall algorithm from ten randomly generating starting points, each of which is a uniformly randomly generated sequence, and retain the best solution obtained over the ten repetitions. 
\end{itemize}
With both \LSONE and \LSTEN, we impose a time limit of 600 seconds on the total run time. In the most extreme case, \LSONE will require 600 seconds, while \LSTEN will require $600 \times 10 = 6000$ seconds. In both variants, the functions $G_1$ and $G_2$ (see Appendix~\ref{subsec:localsearch_feasibility}) and $F$ (see Appendix~\ref{subsec:localsearch_F}) are computed by solving the corresponding MILPs using Gurobi with a single thread. We again implement our procedure in Python and run our experiments on the same Amazon EC2 instance described in Section~\ref{subsec:results_TTPGLOBAL}.

\begin{figure*}
    \centering
    \includegraphics[width=0.8\textwidth]{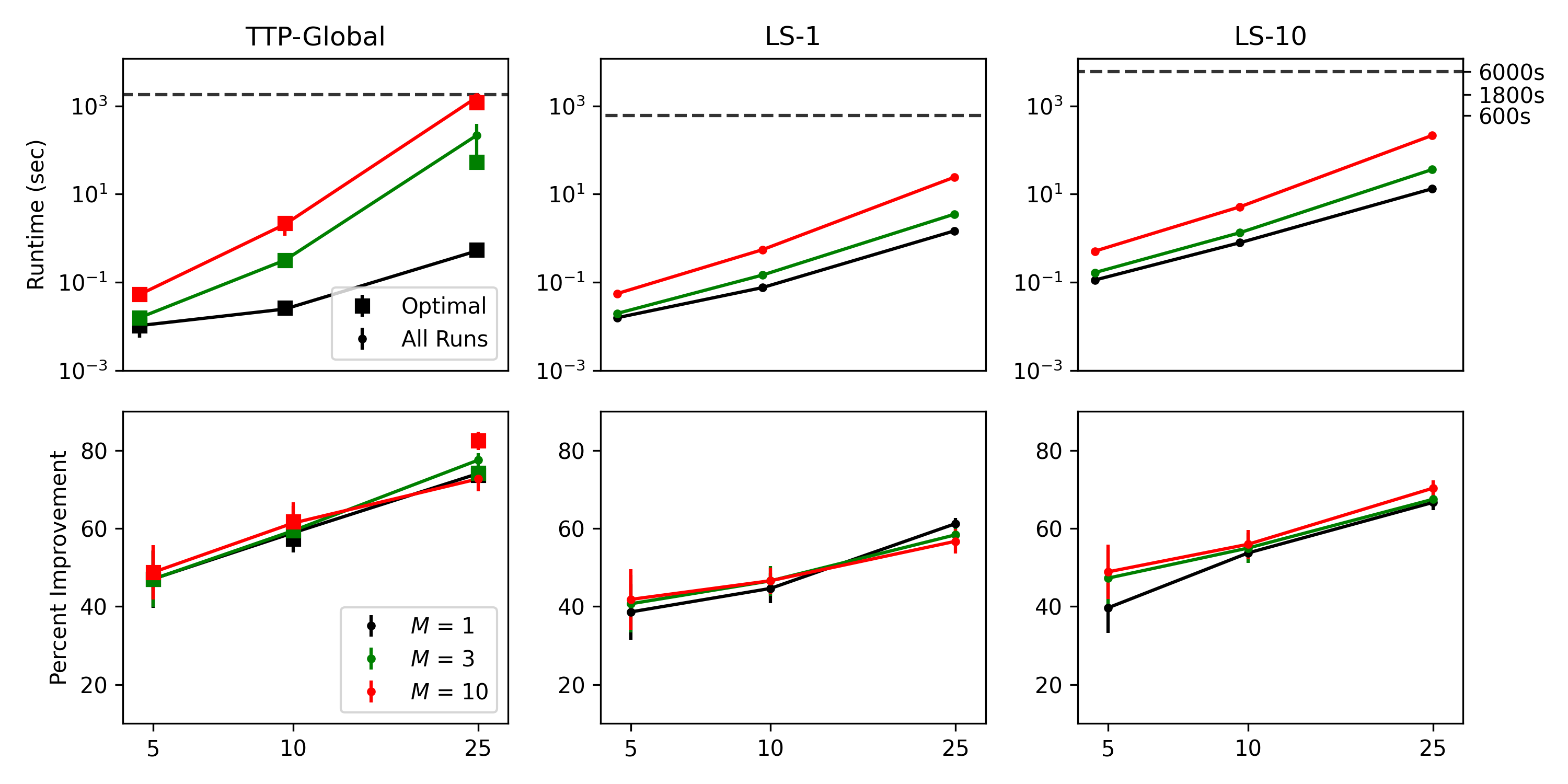}
    \includegraphics[width=0.8\textwidth]{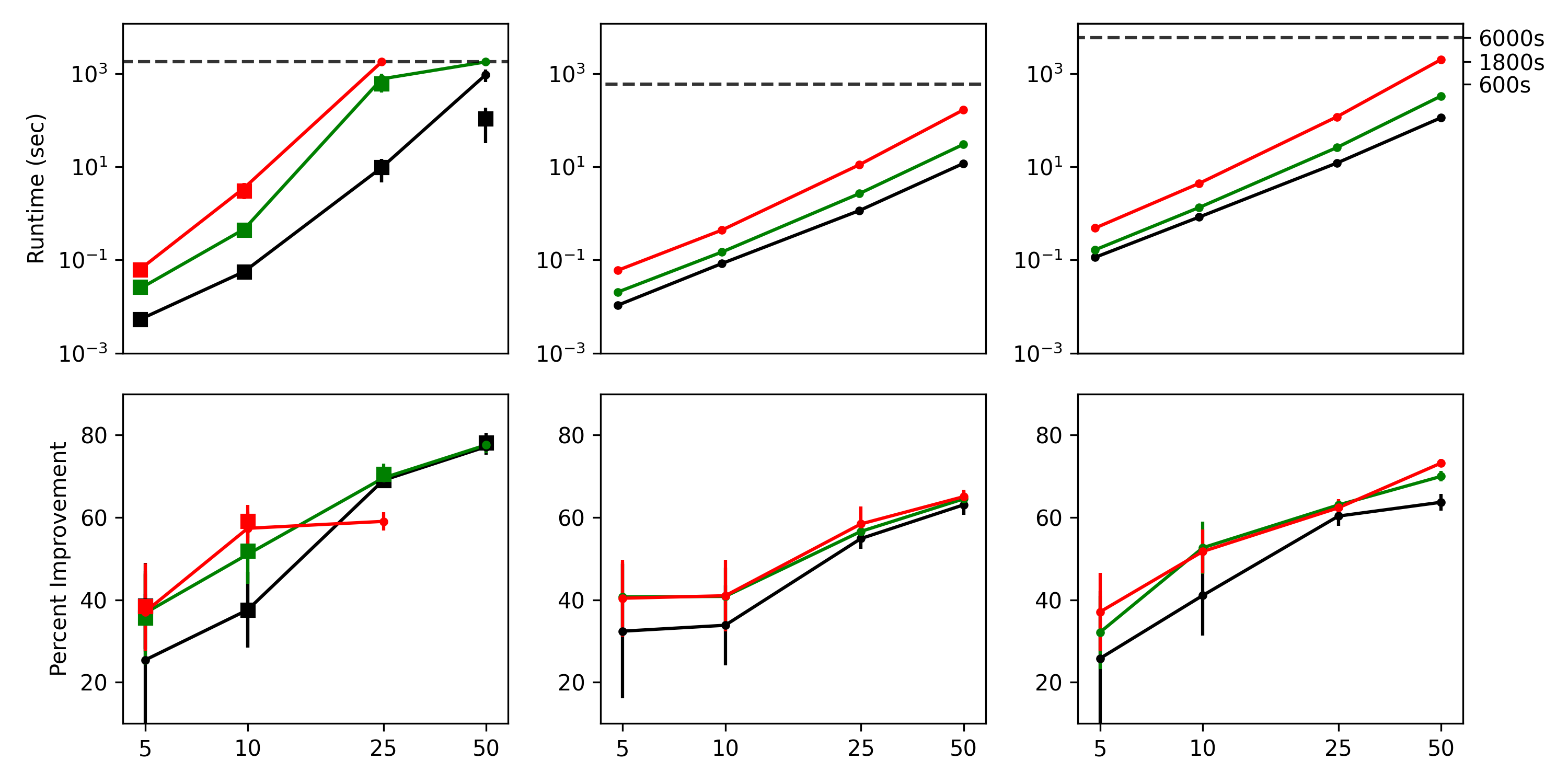}
    \includegraphics[width=0.8\textwidth]{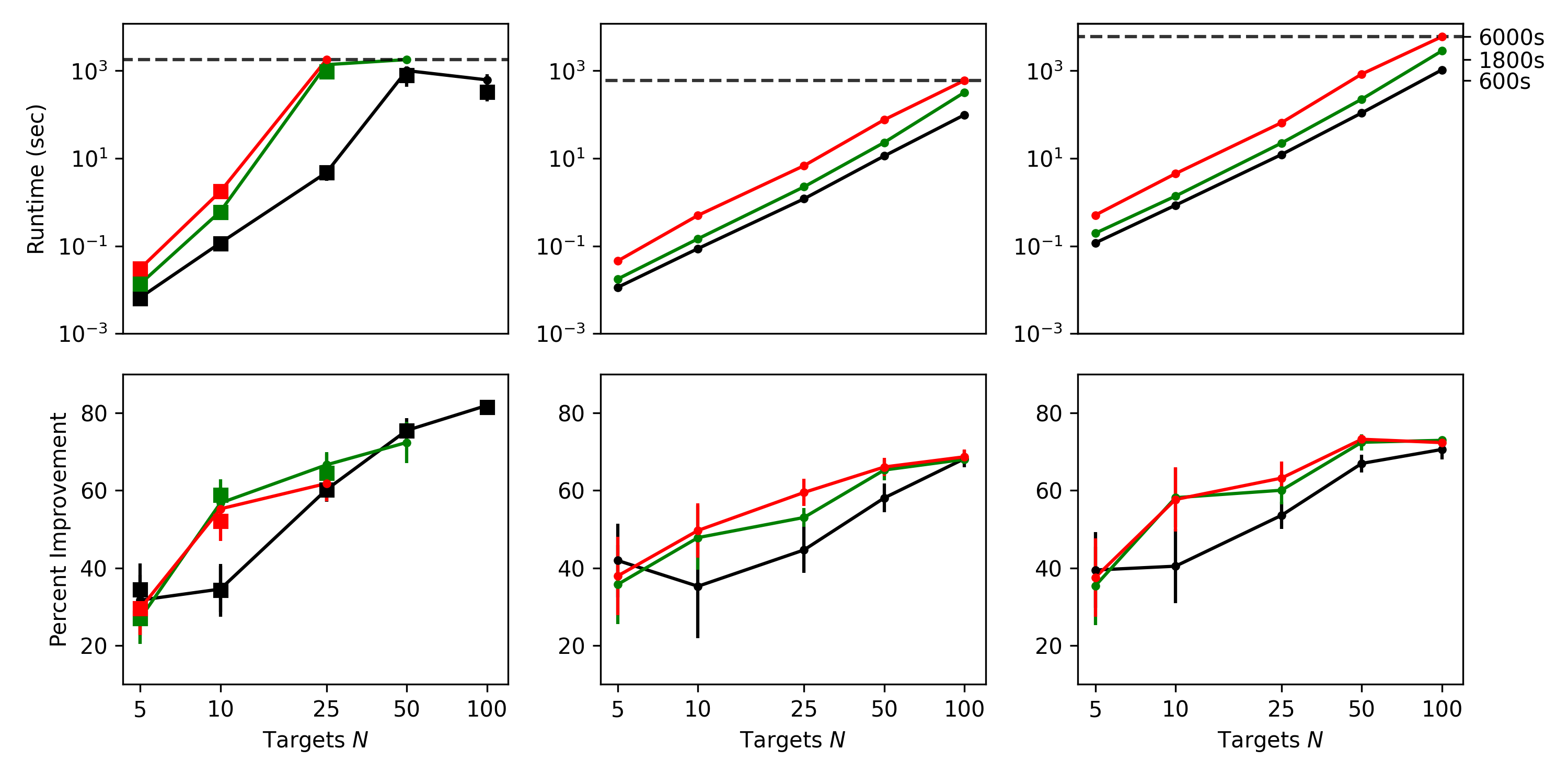}
    \caption{$\Runtime$ and $\RelRed_{\real}$ for each duration type $D$, as found in Tables \ref{table:ttpglobal_results}, \ref{table:LSONE}, and \ref{table:LSTEN}. The first two rows summarize these statistics for the $D=\text{Quarter}$ simulations, the next two for $D=\text{Half}$, and the bottom two for $D=\text{Full}$. The top row in each pair shows the average $\Runtime$ (and optimal subset, for $\TTPGLOBAL$) across all runs for $D$ with varying $M$ as a function of $N$. The bottom row in each pair shows the average $\RelRed_{real}$ for the same values of $D$ and $M$.}
    \label{fig:comparison_full}
\end{figure*}

Table~\ref{table:LSONE} summarizes the results for \LSONE and Figure~\ref{fig:comparison_full} shows run time and slew efficiency for different problem sizes. \LSONE exhibits favorable performance in terms of computation time; in most cases, \LSONE terminates with a locally optimal solution within 600 seconds. The only exception is the $(D, N, M) = (\text{Full}, 100, 10)$ set of instances. Note that \LSONE resulted in a feasible schedule in all but seven of the 360 instances; importantly, \LSONE finds a feasible schedule in all of the instances for parameter combinations for which the \TTPGLOBAL MILP fails (for example, for $(\text{Full}, 100, 10)$, \LSONE produces a feasible schedule in all ten instances in 600 seconds, whereas \TTPGLOBAL fails to find a feasible schedule in all ten instances with 1800 seconds of computation. Of those seven instances in which \LSONE did not find a feasible schedule, six are the same instances which were determined to be infeasible by \TTPGLOBAL, and in one instance, the feasibility procedure (Algorithm~\ref{alg:feasibility}) failed to identify a feasible solution, despite the fact that the instance does admit a feasible solution based on running \TTPGLOBAL. Lastly, in terms of solution quality, the total slew time, as measured by $\SlewTime_{\tau}$ and $\SlewTime_{\real}$, compares favorably to the worst-case bound of $2N$. In the most significant case, with $N = 100$ targets, \LSONE obtains schedules with total slew times that achieve a reduction of approximately 80\% relative to the $2N$ bound.

\begin{table*}
\begin{tabular}{lrrrrrrrr}
\toprule
$D$ & $N$ & $M$ & $\NumFeas$ & $\Runtime$ & $\SlewTime_{\tau}$ & $\RelRed_{\tau}$ & $\SlewTime_{\real}$ & $\RelRed_{\real}$ \\
& & & & (s) & (min) & (\%) & (min) & (\%) \\
\midrule
Quarter & 5 & 1 & 10 & 0.0 & 4.3 & 57.1 & 6.1 & 38.6 \\
Quarter & 5 & 3 & 10 & 0.0 & 4.4 & 56.3 & 5.9 & 40.7 \\
Quarter & 5 & 10 & 10 & 0.1 & 4.1 & 58.7 & 5.8 & 41.8 \\
Quarter & 10 & 1 & 10 & 0.1 & 7.7 & 61.3 & 11.1 & 44.6 \\
Quarter & 10 & 3 & 10 & 0.1 & 7.3 & 63.6 & 10.7 & 46.5 \\
Quarter & 10 & 10 & 10 & 0.6 & 7.1 & 64.7 & 10.7 & 46.6 \\
Quarter & 25 & 1 & 10 & 1.5 & 14.3 & 71.5 & 19.4 & 61.2 \\
Quarter & 25 & 3 & 10 & 3.5 & 13.7 & 72.5 & 20.8 & 58.4 \\
Quarter & 25 & 10 & 10 & 24.3 & 13.9 & 72.2 & 21.7 & 56.7 \\ \midrule
Half & 5 & 1 & 10 & 0.0 & 5.4 & 45.9 & 6.8 & 32.4 \\
Half & 5 & 3 & 10 & 0.0 & 4.8 & 51.8 & 5.9 & 40.7 \\
Half & 5 & 10 & 10 & 0.1 & 4.6 & 53.6 & 6.0 & 40.4 \\
Half & 10 & 1 & 10 & 0.1 & 7.9 & 60.6 & 13.2 & 33.8 \\
Half & 10 & 3 & 10 & 0.1 & 7.7 & 61.5 & 11.8 & 40.8 \\
Half & 10 & 10 & 10 & 0.4 & 7.2 & 64.0 & 11.8 & 41.0 \\
Half & 25 & 1 & 10 & 1.1 & 15.9 & 68.3 & 22.6 & 54.9 \\
Half & 25 & 3 & 10 & 2.7 & 13.1 & 73.8 & 21.7 & 56.6 \\
Half & 25 & 10 & 10 & 11.1 & 13.4 & 73.2 & 20.8 & 58.4 \\
Half & 50 & 1 & 10 & 11.8 & 23.1 & 76.9 & 36.9 & 63.1 \\
Half & 50 & 3 & 10 & 30.4 & 20.3 & 79.7 & 35.5 & 64.5 \\
Half & 50 & 10 & 10 & 167.4 & 22.5 & 77.5 & 35.0 & 65.0 \\ \midrule
Full & 5 & 1 & 9 & 0.0 & 6.2 & 37.6 & 6.6 & 34.4 \\
Full & 5 & 3 & 9 & 0.0 & 5.7 & 43.1 & 7.2 & 27.5 \\
Full & 5 & 10 & 9 & 0.0 & 5.5 & 44.6 & 7.0 & 30.0 \\
Full & 10 & 1 & 8 & 0.1 & 10.0 & 49.9 & 16.4 & 17.9 \\
Full & 10 & 3 & 9 & 0.1 & 9.5 & 52.6 & 11.7 & 41.5 \\
Full & 10 & 10 & 9 & 0.5 & 8.5 & 57.4 & 11.3 & 43.5 \\
Full & 25 & 1 & 10 & 1.2 & 17.5 & 65.0 & 27.7 & 44.7 \\
Full & 25 & 3 & 10 & 2.2 & 17.0 & 66.0 & 23.5 & 53.0 \\
Full & 25 & 10 & 10 & 6.8 & 14.4 & 71.2 & 20.3 & 59.5 \\
Full & 50 & 1 & 10 & 11.3 & 27.6 & 72.4 & 41.9 & 58.1 \\
Full & 50 & 3 & 10 & 23.1 & 23.2 & 76.8 & 34.7 & 65.3 \\
Full & 50 & 10 & 10 & 76.2 & 23.3 & 76.7 & 33.9 & 66.1 \\
Full & 100 & 1 & 10 & 98.7 & 43.7 & 78.2 & 63.6 & 68.2 \\
Full & 100 & 3 & 10 & 321.6 & 37.0 & 81.5 & 63.9 & 68.0 \\
Full & 100 & 10 & 10 & 602.0 & 38.7 & 80.6 & 62.6 & 68.7 \\ \bottomrule
\end{tabular}
\caption{Computational results for \LSONE procedure. \label{table:LSONE} }
\end{table*}

Table~\ref{table:LSTEN} and Figure~\ref{fig:comparison_full} presents analogous results for \LSTEN. 

We found that the \LSTEN schedules were signficantly more efficient than the \LSONE schedules. For example, for $(\text{Full},100,1)$, $\SlewTime_{\tau}$ is 32.6~min for \LSTEN, compared to 43.7~min for \LSONE. 

In cases where the \TTPGLOBAL returned an optimal schedule, this schedule was often much more efficient than \LSTEN and \LSONE. For example, for $(\text{Full},50,1)$, the \TTPGLOBAL MILP was solved to full optimality in seven out of ten instances, and the average $\SlewTime_{\tau}$ was 13.2~min, compared to 27.6~min for \LSONE and 19.9~min for \LSTEN. \LSTEN and \LSONE do not guarantee a globally optimal solution, but gap between local and global optima suggests additional work on heuristic solutions could prove fruitful. 

On the other hand, in cases where the \TTPGLOBAL MILP does not terminate with an optimal solution, it is possible for the local search solution to perform better. For example, for the $(D, N, M) = (\text{Half}, 25, 10)$ experiments, the \LSTEN solution has an average $\SlewTime_{\tau}$ of 10.8~min compared to 12.7~min for \TTPGLOBAL. Lastly, the computation time for \LSTEN is roughly ten times that of \LSONE, as one would expect. However, we note that the ten repetitions are independent, and could be carried out in parallel. This could be attractive from an implementation standpoint, as both \LSONE and \LSTEN were executed with a single-thread, so one could easily execute the local search procedure from multiple starting points in parallel within a multi-threaded computing environment. 

There are several key takeaways from Figure \ref{fig:comparison_full} when $D=\text{Full}$. Adopting static target-to-target slew overheads ($M = 1$), we find \TTPGLOBAL solves the schedule for $N$ up to 100 in most runs. \TTPGLOBAL produces the highest $\RelRed_{\real}$ improvement of above $80\%$ for the $N=100$ case. For smaller cases of $N$, it sees some benefit from higher values of $M$, but cannot find feasible solutions in the time limit for large $N$. The local solvers \LSONE and \LSTEN scale exponentially with $N$, and find local optima for all $N$ in their expected time limits. $\RelRed_{\real}$ benefits from higher $M$ for moderate values of $N$, but also has diminishing returns for $N=100$, and a lower ceiling compared to the global solution. Local search is equipped to find feasible solutions to larger models, but struggles to find solutions near the global optimum at high $N$, regardless of the value of $M$.

\begin{table*}
\begin{tabular}{lrrrrrrrr}
  \toprule
$D$ & $N$ & $M$ & $\NumFeas$ & $\Runtime$ & $\SlewTime_{\tau}$  & $\RelRed_{\tau}$ & $\SlewTime_{\real}$  & $\RelRed_{\real}$  \\ 
& & & & (s) & (min) & (\%) & (min) & (\%) \\
\midrule
Quarter &  5 &  1 & 10 & 0.1 & 4.1 & 59.1 & 6.0 & 39.6 \\ 
  Quarter &  5 &  3 & 10 & 0.2 & 3.7 & 63.3 & 5.3 & 47.2 \\ 
  Quarter &  5 & 10 & 10 & 0.5 & 3.5 & 64.5 & 5.1 & 48.9 \\ 
  Quarter & 10 &  1 & 10 & 0.8 & 5.8 & 70.9 & 9.3 & 53.7 \\ 
  Quarter & 10 &  3 & 10 & 1.3 & 5.8 & 71.2 & 9.0 & 55.0 \\ 
  Quarter & 10 & 10 & 10 & 5.2 & 5.5 & 72.6 & 8.8 & 55.9 \\ 
  Quarter & 25 &  1 & 10 & 13.2 & 10.2 & 79.7 & 16.7 & 66.7 \\ 
  Quarter & 25 &  3 & 10 & 36.1 & 9.9 & 80.2 & 16.3 & 67.5 \\ 
  Quarter & 25 & 10 & 10 & 214.4 & 9.6 & 80.8 & 14.8 & 70.3 \\ \midrule
  Half &  5 &  1 & 10 & 0.1 & 4.9 & 50.8 & 7.4 & 25.8 \\ 
  Half &  5 &  3 & 10 & 0.2 & 4.7 & 53.2 & 6.8 & 32.1 \\ 
  Half &  5 & 10 & 10 & 0.5 & 4.4 & 55.5 & 6.3 & 37.1 \\ 
  Half & 10 &  1 & 10 & 0.8 & 6.6 & 67.2 & 11.8 & 41.1 \\ 
  Half & 10 &  3 & 10 & 1.3 & 6.0 & 70.1 & 9.5 & 52.6 \\ 
  Half & 10 & 10 & 10 & 4.4 & 5.8 & 71.0 & 9.7 & 51.7 \\ 
  Half & 25 &  1 & 10 & 12.1 & 11.2 & 77.7 & 19.8 & 60.3 \\ 
  Half & 25 &  3 & 10 & 26.0 & 10.8 & 78.4 & 18.5 & 63.0 \\ 
  Half & 25 & 10 & 10 & 118.6 & 10.8 & 78.5 & 18.8 & 62.4 \\ 
  Half & 50 &  1 & 10 & 113.1 & 19.1 & 80.9 & 36.3 & 63.7 \\ 
  Half & 50 &  3 & 10 & 330.3 & 16.7 & 83.3 & 30.0 & 70.0 \\ 
  Half & 50 & 10 & 10 & 2003.9 & 17.0 & 83.0 & 26.8 & 73.2 \\ \midrule
  Full &  5 &  1 &  9 & 0.1 & 6.0 & 39.7 & 6.8 & 31.6 \\ 
  Full &  5 &  3 &  9 & 0.2 & 5.3 & 47.0 & 7.3 & 27.1 \\ 
  Full &  5 & 10 &  9 & 0.5 & 5.3 & 47.3 & 7.0 & 29.5 \\ 
  Full & 10 &  1 &  9 & 0.9 & 8.8 & 56.2 & 13.3 & 33.3 \\ 
  Full & 10 &  3 &  9 & 1.4 & 7.7 & 61.7 & 9.4 & 52.9 \\ 
  Full & 10 & 10 &  9 & 4.5 & 6.5 & 67.5 & 9.5 & 52.4 \\ 
  Full & 25 &  1 & 10 & 12.2 & 12.2 & 75.5 & 23.2 & 53.6 \\ 
  Full & 25 &  3 & 10 & 22.3 & 12.1 & 75.9 & 20.0 & 60.1 \\ 
  Full & 25 & 10 & 10 & 65.4 & 11.4 & 77.1 & 18.4 & 63.2 \\ 
  Full & 50 &  1 & 10 & 109.2 & 19.9 & 80.1 & 33.0 & 67.0 \\ 
  Full & 50 &  3 & 10 & 224.9 & 19.5 & 80.5 & 27.5 & 72.5 \\ 
  Full & 50 & 10 & 10 & 842.0 & 17.9 & 82.1 & 26.8 & 73.2 \\ 
  Full & 100 &  1 & 10 & 1052.1 & 32.6 & 83.7 & 58.8 & 70.6 \\ 
  Full & 100 &  3 & 10 & 2852.0 & 29.8 & 85.1 & 54.1 & 72.9 \\ 
  Full & 100 & 10 & 10 & 6025.6 & 32.3 & 83.8 & 55.3 & 72.4 \\   \bottomrule
\end{tabular}
\caption{Computational results for \LSTEN procedure. \label{table:LSTEN} }
\end{table*}

% Lastly, we remark that this overall procedure, which seeks to obtain a feasible solution to a constrained optimization problem by relaxing a set of constraints and then minimizing the sum of violations of these constraints, is common in the optimization literature. In particular, in linear programming, the two-phase method (see CITEBERTSIMASTISTSIKLISHERE) is an algorithm for solving linear programs that involves a first phase in which 


\newpage
\bibliography{PASPsample631}{}
\bibliographystyle{aasjournal}



\end{document}